\documentclass[10pt,aps,prb,twocolumn,preprintnumbers,amsmath,amssymb,floatfix,showpacs,citeautoscript,superscriptaddress]{revtex4-2} %
\usepackage[latin1]{inputenc}
\usepackage[T1]{fontenc}
\usepackage[english]{babel}
\usepackage[pdftex]{graphicx}
\usepackage{amsmath}
\usepackage{times}
\usepackage[usenames,dvipsnames,svgnames,table]{xcolor}
\usepackage[colorlinks,plainpages=false,linkcolor=blue,urlcolor=blue,citecolor=blue,pdfpagemode=UseNone]{hyperref}
\usepackage{hyperref}

\newcommand{\Tc}{T_{c}}
\newcommand{\lambdaL}{\lambda_{\mathrm{L}}}

\begin{document}

\title
{
\boldmath
Towards the discovery of high critical magnetic field superconductors
}

\author{Benjamin Geisler}
\email{benjamin.geisler@ufl.edu}
\affiliation{Department of Physics, University of Florida, Gainesville, Florida 32611, USA}
\affiliation{Department of Materials Science and Engineering, University of Florida, Gainesville, Florida 32611, USA}
\author{Philip M. Dee}
\affiliation{Computational Sciences and Engineering Division, Oak Ridge National Laboratory, Oak Ridge, Tennessee, 37831-6494, USA\looseness=-1}
\author{James J. Hamlin}
\affiliation{Department of Physics, University of Florida, Gainesville, Florida 32611, USA}
\author{Gregory R. Stewart}
\affiliation{Department of Physics, University of Florida, Gainesville, Florida 32611, USA}
\author{Richard G. Hennig}
\affiliation{Department of Materials Science and Engineering, University of Florida, Gainesville, Florida 32611, USA}
\affiliation{Quantum Theory Project, University of Florida, Gainesville, Florida 32611, USA}
\author{P.J. Hirschfeld}
\affiliation{Department of Physics, University of Florida, Gainesville, Florida 32611, USA}

\date{\today}

\begin{abstract}
Superconducting materials are of significant technological relevance for a broad range of applications, and intense research efforts aim at enhancing the critical temperature $T_{c}$. Intriguingly, while numerous studies have explored different computational and machine-learning routes to predict $T_{c}$, the fundamental role of the critical magnetic field has so far been overlooked.
Here we open a new frontier in superconductor discovery by presenting a consistent computational database of critical fields $H_{c}$, $H_{c1}$, and $H_{c2}$ for over $7{,}300$ electron-phonon-paired superconductors covering distinct materials classes.
A theoretical framework is developed that combines $\alpha^2F(\omega)$ spectral functions and highly accurate Fermi surfaces from density functional theory with clean-limit Eliashberg theory to obtain the coherence lengths, London penetration depths, and Ginzburg-Landau parameters.
We discover an unexpectedly large number of Type-I superconductors and
show that larger unit cells generically support higher critical fields and Type-II behavior.
We identify the importance of going beyond BCS theory by including strong-coupling corrections to the superconducting gap and electron-phonon renormalizations of the effective mass for predictions of critical fields across materials.
These results provide a framework for foundational AI models that realize the concept of inverse materials design for high-$T_{c}$ and high-critical-field superconductors.
\end{abstract}

\maketitle

\section{Introduction}

The technological relevance of superconducting materials has driven substantial efforts to identify next-generation compounds with enhanced critical temperature $\Tc$ \cite{Bednorz1986,Wu1987,Schilling1993,Nagamatsu2001,Kamihara2008,StewartRMP2011,Cao2018,Kamysbayev2020,FLORESLIVAS2020,ROADMAP,Molodyk2023,PellegriniSanna:2024,Gao2025}.
A successful route towards this goal is the combination of computational and theoretical methods for materials discovery.
This strategy has led to notable breakthroughs, most prominently the prediction and subsequent experimental confirmation of high-$\Tc$ hydride superconductors under pressure \cite{Duan14, Drozdov_H3S_2015,HemleyLaH10,Eremets_hydride_review2019,PickettRMP}.
The recent rise of machine learning has considerably accelerated these explorations \cite{Stanev2018,Meredig2018,Hutcheon2020,Xie2022,ZhangPickard:2022,Choudhary2022,Wines2023,Sommer2022_3DSC,betenet, beenet, prakash-guideddiffusion},
a necessity in light of the immense search space for superconductor discovery.
However, superconductivity as a macroscopic quantum many-body phenomenon remains extremely challenging to predict.

It has become increasingly clear that $\Tc$ alone is not the decisive metric of technological usefulness.
Critical magnetic fields and the associated critical currents also play a central role, as do mechanical aspects such as ductility and processability \cite{Larbalestier2001}.
For example, high-$\Tc$ cuprates are limited by their layered structure and resulting two-dimensional superconductivity, which facilitate vortex-liquid formation \cite{Vortices:1994}.
Similarly, MgB$_2$, one of the highest-$\Tc$ conventional superconductors, faces manufacturing challenges and exhibits a two-gap superconducting structure that, while being a feature of strong scientific interest \cite{Choi2002, Margine2013}, further limits practical applications compared to e.g.\ Nb-based compounds.

Interestingly, the technological importance of critical magnetic fields has been largely overlooked:
High critical fields are essential for many key applications,
including high-performance magnets for high-field laboratories, fusion reactors, particle accelerators, and MRI machines.

These considerations require a paradigm shift in the field.
The primary goal is to avoid the need for helium-based cooling, as the cost of cryogenic operation increases by orders of magnitude as $\Tc$ decreases.
Thus, we underline the following three primary requirements for a practical superconductor:
(i)~$\Tc$ sufficiently high to allow a safe operating temperature well above liquid helium, e.g., $\gtrsim 20$~K;
(ii)~Critical fields to achieve required critical currents and resulting magnetic fields for applications, e.g., $\gtrsim 10$~T \cite{NAS_HighField2024};
(iii)~Ductility and processability to enable wire fabrication.
Hence, rather than focusing solely on high $\Tc$, it is imperative to systematically identify superconductors with large critical fields.

\begin{figure*}
\begin{center}
\includegraphics[width=\linewidth]{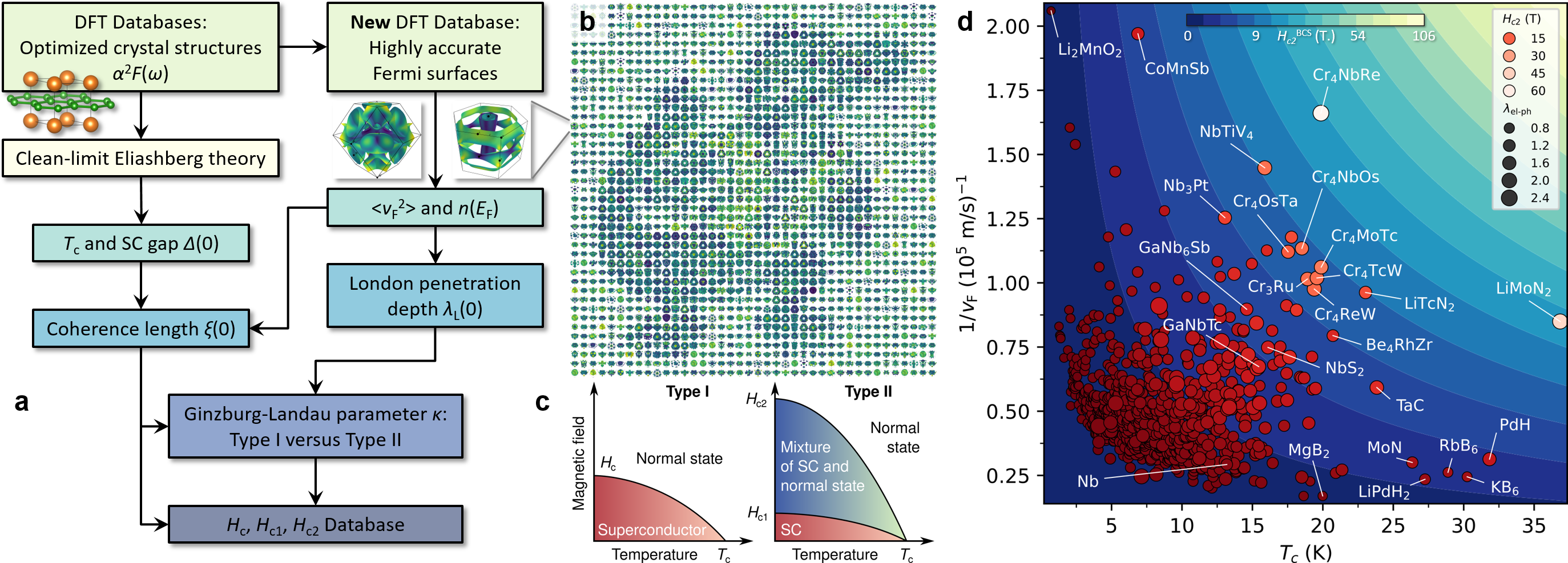}
\caption{\label{fig:Overview}
\textbf{\boldmath First-principles database of superconducting critical magnetic fields.}
\textbf{a}~The workflow developed here to obtain a consistent database of critical fields and related superconducting properties entirely from first principles.
\textbf{b}~The fundamental ingredient is a new database of Fermi surfaces, colored here by the Fermi velocities; a selection has been compiled in the form of a mosaic where each pixel is a Fermi surface.
\textbf{c}~Explanation of the different critical fields.
\textbf{d}~Upper critical fields at zero temperature for all Type-II superconductors in our database, systematically ranked in a $(\Tc, 1/v_\text{F})$ scatterplot.
The blue-yellow background shows $H_{c2}^\text{BCS}$ to structure the data, as described in the text.
The point colors reflect the actual upper critical fields $H_{c2} > H_{c2}^\text{BCS}$ determined by using the calculated Eliashberg gaps,
which include strong-coupling corrections,
and considering effective-mass renormalizations of the Fermi velocities due to electron-phonon interactions $\lambda_\text{el-ph}$.
}
\end{center}
\end{figure*}

Here we take a first step in this new direction by establishing a first-principles workflow to predict thermodynamic critical fields $H_{c}$, lower critical fields $H_{c1}$, and upper critical fields $H_{c2}$ that combines clean-limit Eliashberg theory with $\alpha^2F(\omega)$ spectral functions and Fermi surface properties from density functional theory.
We construct a comprehensive database for more than $7{,}300$ electron-phonon-paired superconductors spanning diverse materials classes.
Using this unique perspective, we identify compounds with high critical fields and uncover surprising trends that challenge prevailing wisdom in the field.
In particular, we find an unexpectedly large number of Type-I superconductors and show that higher critical fields and \mbox{Type-II} behavior are favored in materials with larger unit cells.
We further demonstrate the necessity of going beyond BCS theory to achieve reliable predictions of critical fields.
These results lay the groundwork for future foundational AI models enabling inverse materials design of high-$T_{c}$ and high-critical-field superconductors.

\section{Results}

\subsection{Workflow: First-principles database of superconducting critical magnetic fields}

We construct a database of superconducting critical fields for $\sim 7{,}300$ realistic materials
by combining density functional theory (DFT) \cite{KoSh65} and clean-limit Eliashberg theory \cite{EliashbergInteractionBetweenElAndLatticeVibrInASC1960, Carbotte1990RMP}.
All quantities are obtained consistently from first principles, %
using a workflow we developed and describe below [Fig.~\ref{fig:Overview}(a)].
Importantly, our philosophy here is to present and discuss a broad set of materials,
prioritizing comprehensive coverage. Therefore, we deliberately avoid excluding any compounds based on
physical intuition, and retain all materials in the final results, including outliers.

We start from two established databases of electron-phonon superconductors that provide
DFT-optimized crystal structures and corresponding calculated $\alpha^2F(\omega)$ Eliashberg spectral functions [Fig.~\ref{fig:Overview}(a)],
from our group \cite{betenet} and from Cerqueira \textit{et al.} \cite{Cerqueira2023},
complemented by new calculations for this work. %
These parent databases cover a wide variety of materials across the periodic table.

By solving the isotropic Eliashberg equations \cite{EPW-1, EPW-2} %
in a high-throughput framework,
we explicitly calculate the superconducting critical temperature $\Tc$
and the superconducting gap at zero temperature $\Delta(0)$ for the reference isotropic system,
explicitly accounting for strong-coupling corrections.
In parallel, we perform high-throughput DFT calculations \cite{Kresse1996, Kresse1996b, Kresse1999} %
to obtain accurate Fermi surfaces for all materials [Fig.~\ref{fig:Overview}(b)]. %
Subsequently,
we calculate a key ingredient, the averaged Fermi velocities:
\begin{equation}
\langle v_\text{F}^2 \rangle = \frac{1}{A_\text{FS} \, \hbar^2} \int_\text{FS} dn \, \left( \nabla_{\vec{k}} \, \varepsilon(\vec{k}) \right)^2 = \langle v_x^2 \rangle + \langle v_y^2 \rangle + \langle v_z^2 \rangle .
\end{equation}
Our methodology provides the sheet- and direction-resolved components of the squared velocity operator
(i.e., the anisotropic information).
Here, we use the isotropic average, integrated over the entire Fermi surface. %
The DFT high-throughput sampling also provides accurate total densities of states $n(E_\text{F})$ by using the tetrahedron method, which we normalize by the unit cell volume.
Further numerical details are reported in the Methods section.

From these fundamental quantities, we calculate the superconducting coherence length,
\begin{equation}
\xi(0) = \frac{\hbar \, \sqrt{ \langle v_\text{F}^2 \rangle }}{\pi \, \langle\Delta(0)\rangle},
\end{equation}
and the London penetration depth, %
\begin{equation}
\label{eq:LL}
\lambdaL(0) = \sqrt{\frac{m}{\mu_0 n_\text{s} e^2}},
\end{equation}
where $\langle\Delta(0)\rangle$ is an appropriately averaged gap function over the Fermi surface \cite{Whitmore1981}, which we replace by $\Delta(0)$ in the absence of information on the anisotropy of the pairing interaction.
The expressions given apply formally to single-band systems in the BCS clean limit as $T \to 0$ \cite{Tinkham-SC}.
While Eq.~(\ref{eq:LL}) is often used to infer the superfluid density $n_\text{s}$ from a measured $\lambdaL$,
here we recast it into a form in which all quantities are obtained consistently from DFT:
\begin{equation}
\lambdaL(0) = \sqrt{ \frac{3}{\mu_0 \, e^2 \, n(E_\text{F}) \, \langle v_\text{F}^2 \rangle } },
\end{equation}
by substituting the effective mass and superfluid density with the volume-normalized total density of states and averaged isotropic Fermi velocity.
Experimentally, the measured superconducting penetration depths are always larger than the BCS $\lambdaL(0)$ \cite{Tinkham-SC}.

The BCS expressions for $\xi$ and $\lambdaL$ are renormalized within the more complete
Eliashberg theory of superconductivity by the electron-phonon coupling strength,
\begin{equation}
\lambda_\text{el-ph} = 2 \int d\omega \, \alpha^2 F(\omega)/\omega
\end{equation}
via the factors
\begin{equation}
    \lambdaL \rightarrow \lambdaL \sqrt{1+\lambda_\text{el-ph}},
    \quad
    \xi \rightarrow \xi / (1+\lambda_\text{el-ph}) ,
\end{equation}
which follows from
\begin{equation}
    n(E_\text{F}) \rightarrow n(E_\text{F})(1+\lambda_\text{el-ph}) ,
    \quad
    v_\text{F} \rightarrow v_\text{F}/(1+\lambda_\text{el-ph}) ,
\end{equation}
accounting for the effective-mass dressing of the electron by its interaction with the phonon system.
We find that these renormalizations lead to substantial modifications of
the Ginzburg-Landau parameter $\kappa$ and the corresponding critical fields,
as detailed below.

The Ginzburg-Landau parameter, defined as
\begin{equation}
\kappa = \lambdaL / \xi,
\end{equation}
allows us to fundamentally distinguish between Type-I \mbox{($\kappa < 1/\sqrt{2}$)}
and Type-II superconductors \mbox{($\kappa > 1/\sqrt{2}$)}.

The upper critical field of Type-II superconductors at zero temperature is given by
\begin{equation}
\label{eq:Hc2}
H_{c2} = \frac{\Phi_0}{2 \pi \xi^2}
= \frac{\pi^2}{2 e \hbar} \frac{\Delta^2}{ \langle v_\text{F}^2 \rangle },
\end{equation}
and depends essentially on the superconducting gap and the averaged Fermi velocity,
with the superconducting flux quantum $\Phi_0 = h/2e$.

For the lower critical field of Type-II superconductors, we employ the expression proposed by Brandt~\cite{Brandt-Hc1:2011}:
\begin{equation}
\label{eq:Hc1}
H_{c1} = \frac{\Phi_0}{4 \pi} \frac{\ln(\kappa) + \alpha(\kappa)}{\lambdaL^2}
= H_{c2} \frac{\ln(\kappa) + \alpha(\kappa)}{2 \kappa^2},
\end{equation}
where $\alpha(\kappa)$ has been determined numerically. %
Interestingly, the right-hand side shows that
$H_{c1}$ can be expressed in terms of $H_{c2}$ multiplied by a $\kappa$-dependent factor.

The thermodynamic critical field of Type-I superconductors follows from the Ginzburg-Landau expression:
\begin{equation}
\label{eq:Hc}
H_{c} = \frac{\Phi_0}{2 \sqrt{2} \pi} \frac{1}{\lambdaL \, \xi}
= H_{c2} \frac{1}{\sqrt{2} \kappa}.
\end{equation}
Again, $H_{c}$ can be expressed in terms of (a hypothetical) $H_{c2}$ divided by $\sqrt{2}\kappa$.
These equations emphasize that all critical fields depend on  $\xi$, %
whereas $H_{c}$ and $H_{c1}$ are in addition functions of $\lambdaL$. %
The three critical fields, expressed as functions of the Ginzburg-Landau parameter, have to coincide at $\kappa = 1/\sqrt{2}$;
this important physical condition is satisfied by the present approach.

\begin{figure*}
\begin{center}
\includegraphics[width=\linewidth]{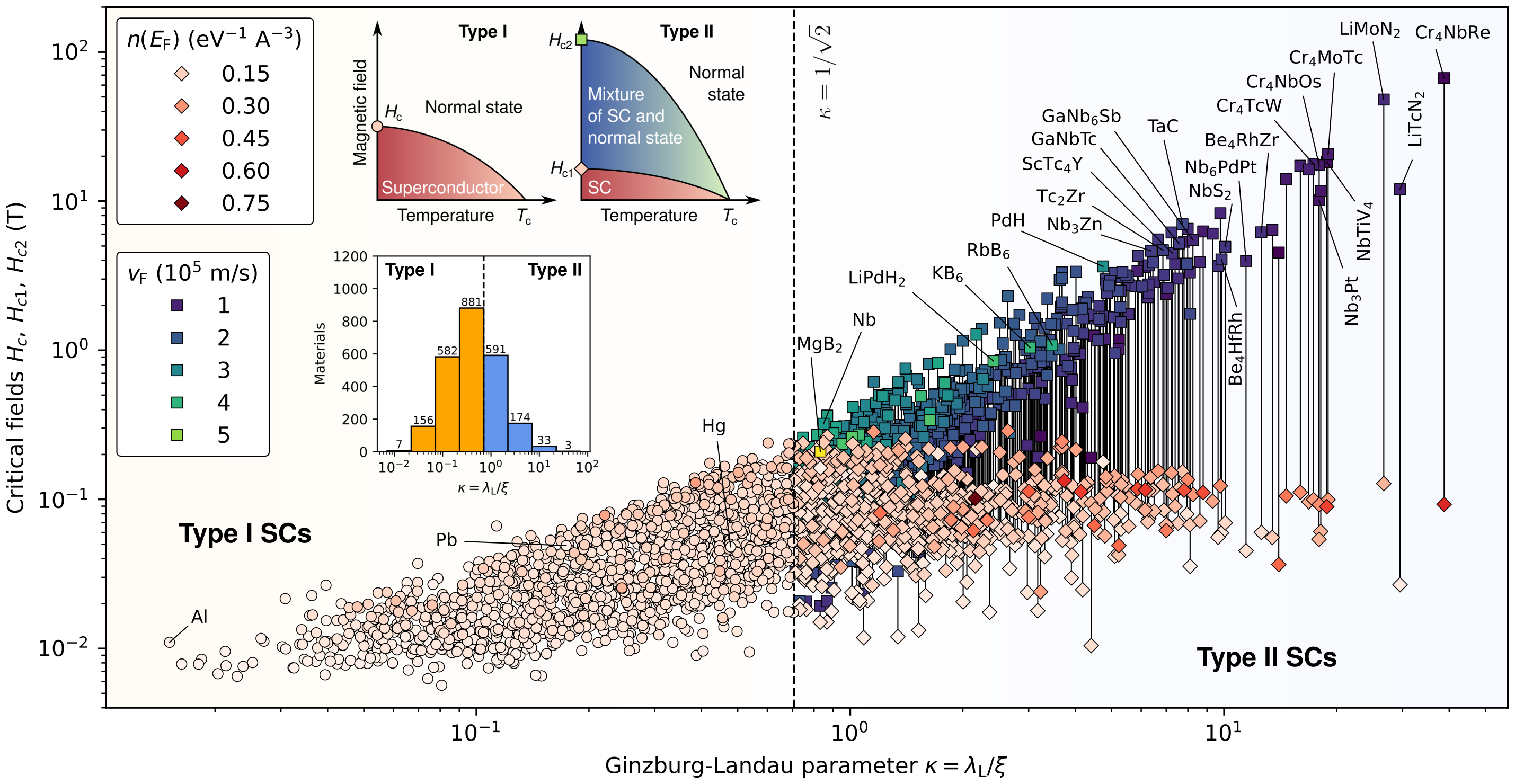}
\caption{\label{fig:HcKappa}
\textbf{\boldmath Critical fields as a function of the Ginzburg-Landau parameter~$\kappa$.}
Type-I critical fields (left) are represented by a single point per material.
For Type-II superconductors (right), upper and lower critical fields are represented by a square-diamond pair, respectively, connected by a vertical line.
Red colors encode the electron-phonon renormalized density of states at the Fermi level $n(E_\text{F})$, with is particularly relevant for $H_{c}$ and $H_{c1}$.
Green-purple colors represent the electron-phonon renormalized Fermi velocity $v_\text{F}$. %
Selected materials are labeled for reference.
The histogram visualizes the distribution of $\kappa$ and thus the number of \mbox{Type-I} versus \mbox{Type-II} superconductors.
The figure focuses on materials with $\Tc>1$~K. All properties have been predicted consistently from DFT.
}
\end{center}
\end{figure*}

\subsection{Exploration of the superconducting critical field data}

Figure~\ref{fig:Overview}(d) shows the upper critical fields $H_{c2}$, which are of key technological importance, for all materials
that are predicted as Type-II superconductors in our database, systematically ranked in a $(\Tc, 1/v_\text{F})$ scatter plot.
The blue-yellow background helps to structure the data by visualizing a BCS approximation for the upper critical fields:
\begin{equation}
H_{c2}^\text{BCS}(\Tc, v_\text{F}) = \frac{\pi^2}{2 e \hbar} \frac{(1.764 \, k_\text{B} \Tc)^2}{ v_\text{F}^2 } .
\end{equation}
In contrast, the point colors reflect the actual upper critical fields $H_{c2}$ based on the Eliashberg gaps,
which exceed $H_{c2}^\text{BCS}$ due to the strong-coupling corrections discussed in more detail below. %
The plot shows how a material may attain a high $H_{c2}$:
Either by exhibiting a high $\Tc$ (i.e., large $\Delta$) or a low Fermi velocity, which both influence $H_{c2}$ quadratically via the coherence length $\xi$.
Interestingly, we find that some materials successfully balance both quantities,
compensating for a slightly lower $\Tc$ by a higher inverse Fermi velocity.
It can also be seen that the higher $H_{c2}$ ranks are increasingly sparsely populated.
Notably, several putative higher-$\Tc$ materials in the dataset, such as PdH, KB$_6$, RbB$_6$ and LiPdH$_2$, are \textit{not} among the top-$H_{c2}$ compounds.

The consistent dataset permits a series of new and unique perspectives on the magnetic response of conventional superconductors.
In Fig.~\ref{fig:HcKappa}, we plot $H_{c}$, $H_{c1}$, and $H_{c2}$ as a function of the predicted Ginzburg-Landau parameter $\kappa$,
focusing on materials with $\Tc > 1$~K.
We find that the critical fields span four orders of magnitude, considerably exceeding the range of the corresponding $\Tc$ ($1$-$37$~K).
At the boundary between \mbox{Type-I} and Type-II superconductivity ($\kappa = 1/\sqrt{2}$), one can see how $H_{c1}$ and $H_{c2}$ merge into $H_{c}$.
We find that \mbox{Type-II} superconductors reach critical fields that exceed those of \mbox{Type-I} materials by several orders of magnitude
(note the logarithmic scale):
While Type-I superconductors attain at most $H_{c} \sim 0.24$~T here,
upper critical fields of $H_{c2}>66$~T can be observed among the Type-II superconductors.
Concomitantly, $H_{c1}$ ranges between $\sim 10^{-2}$ and $3\times 10^{-1}$~T.

We observe that all high-$H_{c2}$ materials are characterized by a relatively low average Fermi velocity $v_\text{F}$ (represented by dark-purple squares).
Notably, although $1/v_\text{F}$ correlates with the density of states at the Fermi energy $n(E_\text{F})$ (red diamonds)
with a Pearson coefficient of 0.66,
there are several counterexamples such as LiTcN$_2$.
Hence, $v_\text{F}$ and $n(E_\text{F})$ should be considered as individual quantities.
Intriguingly, this implies that a high $n(E_\text{F})$, while often beneficial for $\Tc$, is \textit{not} a sufficient indicator for a high critical magnetic field. %

\begin{figure*}
\begin{center}
\includegraphics[width=\linewidth]{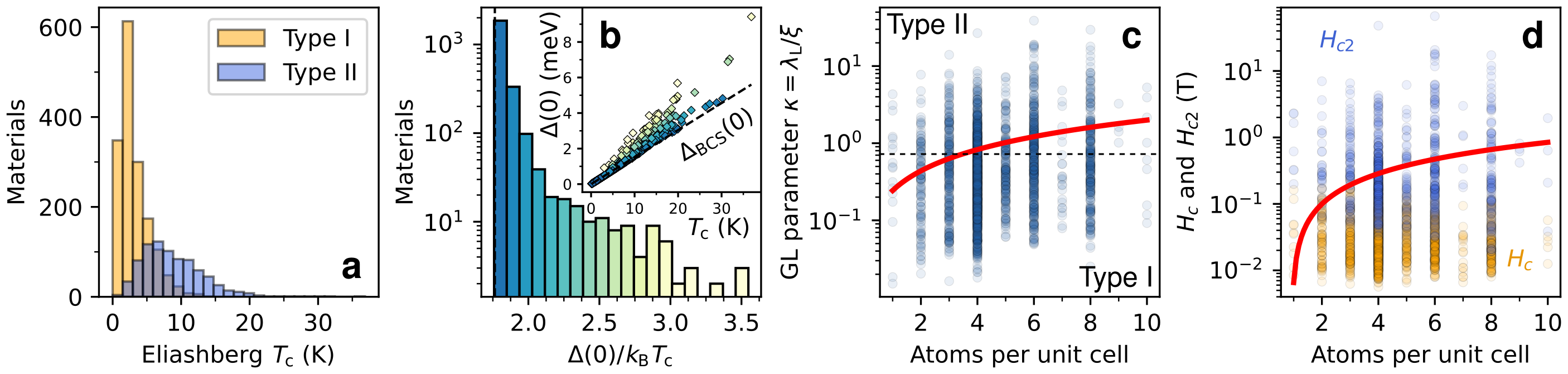}
\caption{\label{fig:Insights}
\textbf{\boldmath Statistical analysis of the superconducting properties.}
\textbf{a}~Distribution of $\Tc$ for \mbox{Type-I} versus \mbox{Type-II} superconductors.
\textbf{b}~The distribution of $\Delta(0) / k_\text{B} \Tc$ shows a clear maximum near the BCS reference, yet concomitantly a substantially extended tail reaching up to twice that value. The inset displays the Eliashberg superconducting gaps (diamonds) versus $\Tc$, colored by $\Delta(0) / k_\text{B} \Tc$.
Comparison with the BCS expression (dashed line) highlights the strong-coupling corrections to the gaps.
\textbf{c}~Correlation of the Ginzburg-Landau parameter $\kappa$ with the unit-cell size.
\textbf{d}~Correlation of the combined $H_c$ (orange; Type~I) and $H_{c2}$ (blue; Type~II) data with the unit-cell size. The linear regression lines are shown in red.
Panels \textbf{a}, \textbf{c}, and \textbf{d} focus on materials with $\Tc>1$~K.
}
\end{center}
\end{figure*}

Figure~\ref{fig:Insights}(a) shows that Type-II superconductors exhibit, in addition to higher critical fields, a general trend towards higher $\Tc$ than Type-I superconductors.
However, neither a large $\Tc$ nor a strong electron-phonon coupling $\lambda$ are a guarantee for a high critical field [Fig.~\ref{fig:Overview}(d)].
This demonstrates that the critical magnetic fields are considerably more complex quantities than the critical temperatures,
which renders them even more challenging to predict.

We find that the strong-coupling corrections to the superconducting gap as obtained from Eliashberg theory significantly enhance the predicted critical fields over a simple BCS approach.
To illustrate this, the inset of Fig.~\ref{fig:Insights}(b) compares the Eliashberg gaps to the weak-coupling limit:
\begin{equation}
\Delta_\text{BCS}(0) = 1.764 \, k_\text{B} \Tc .
\end{equation}
For the paradigmatic strong-coupling superconductor Pb, we obtain a $+17\%$ increase of the gap,
in good agreement with the literature \cite{Mitrovic1984},
and thus of $H_c\sim \Delta$.
For Type-II superconductors, this effect is even more pronounced
and $H_{c2} \sim \Delta^2$ is enhanced by $+37\%$ for Nb and by $+131\%$ for La$_3$Tl.
Intriguingly, the distribution of gap ratios $\Delta(0) / k_\text{B} \Tc$
presents a clear maximum at the BCS reference %
and simultaneously shows a substantially extended tail with exponential decay reaching up to 3.55,
which is more than twice the BCS value. %
To our knowledge, this is the first time that such a systematic statistical analysis of the Eliashberg strong-coupling corrections is discussed for a large set of realistic materials.

It is a common belief that there exist far more Type-II than Type-I superconductors.
This prevailing impression in the community may be strongly shaped by high-$T_c$ and unconventional superconductors, such as cuprates, Fe-based superconductors, and heavy-fermion systems.
On the other hand, Roberts reported already in 1976 far more Type-II than \mbox{Type-I} superconductors \cite{Roberts-SC:1976} -- long before the advent of unconventional superconductivity.
Surprisingly, our findings stand in sharp contrast to this expectation:
We observe a larger number of Type-I than Type-II superconductors in our dataset,
even if we focus exclusively on compounds with $\Tc>1$~K (1626 versus 801 materials; Fig.~\ref{fig:HcKappa}).
We deduce several insights from this observation.
First, our analysis pertains to the clean limit:
Impurities and microstructural disorder, ubiquitous in real materials,
tend to reduce the coherence length $\xi$ and increase the London penetration depth $\lambda_L$,
thereby driving nominally Type-I compounds towards Type-II behavior.
The observed predominance of Type-I superconductivity thus emerges naturally within a clean-limit framework and demonstrates that our approach has the potential to uncover heretofore unrecognized systematic trends in superconducting materials.

Second, we examine the additional possibility that the surprising number of Type-I materials may be related to the restriction of our database to materials with smaller unit cells.
In Fig.~\ref{fig:Insights}(c), we analyze whether the predicted Ginzburg-Landau parameter $\kappa$ correlates with the number of atoms per unit cell, with the hypothesis that larger, more complex crystal structures may generically support higher critical fields.
Intriguingly, the red linear regression line shows a clear positive trend,
spanning an order of magnitude from $\kappa=0.2$ to $2$ as the number of atoms increases from 1 to~10.
Figure~\ref{fig:Insights}(d) uncovers a similar correlation directly for the critical fields $H_{c}$ and $H_{c2}$.
These results provide a statistical confirmation of our hypothesis for a dataset of realistic materials
and suggest that extending it to larger crystal structures in the future will increase the number of Type-II superconductors significantly.
These fundamental insights are particularly relevant for identifying novel high-performance superconductors with small unit cells and a three-dimensional electronic structure.

\begin{table}[b]
\centering
\caption{\label{tab:TopHc2NoSpin}Superconducting parameters (predicted) for the top-$H_{c2}$ Type-II compounds that do not include Cr, Mn, V, and Ti.}
\footnotesize
\begin{ruledtabular}
\begin{tabular}{l c c c c c c c}
Material & Spcgrp. & $\Tc^{\mathrm{El}}$~(K) & $\lambdaL$~(nm) & $\xi$~(nm) & $\kappa$ & $H_{c1}$~(T) & $H_{c2}$~(T) \\
\hline
LiMoN$_2$     & R3m          & 36.8 & 70.0 & 2.62 & 26.7 & 0.128 & 48.0 \\
LiTcN$_2$     & I$\bar{4}$2d & 23.0 & 154.7 & 5.24 & 29.5 & 0.027 & 12.0 \\
Nb$_3$Pt      & Pm$\bar{3}$n & 13.0 & 102.0 & 5.69 & 17.9 & 0.054 & 10.2 \\
TaC           & P$\bar{6}$m2 & 23.8 & 53.1 & 6.86 & 7.7 & 0.153 & 7.0 \\
Be$_4$RhZr    & F$\bar{4}$3m & 20.7 & 91.8 & 7.30 & 12.6 & 0.060 & 6.2 \\
GaNb$_6$Sb    & Pm$\bar{3}$  & 14.6 & 63.9 & 7.75 & 8.3 & 0.108 & 5.5 \\
GaNbTc        & I4mm         & 15.4 & 60.2 & 7.96 & 7.6 & 0.118 & 5.2 \\
NbS$_2$       & P6$_3$/mmc   & 16.1 & 82.2 & 8.16 & 10.1 & 0.070 & 4.9 \\
Tc$_2$Zr      & Fd$\bar{3}$m & 15.3 & 57.5 & 8.38 & 6.9 & 0.125 & 4.7 \\
Nb$_3$Zn      & Pm$\bar{3}$n & 15.1 & 53.5 & 8.41 & 6.4 & 0.141 & 4.6 \\
ScTc$_4$Y     & F$\bar{4}$3m & 12.8 & 62.4 & 8.58 & 7.3 & 0.109 & 4.5 \\
Be$_4$HfRh    & F$\bar{4}$3m & 19.2 & 88.6 & 9.00 & 9.8 & 0.060 & 4.1 \\
Nb$_6$PdPt    & Pm$\bar{3}$  & 12.7 & 104.2 & 9.11 & 11.4 & 0.045 & 4.0 \\
La$_3$Tl      & Pm$\bar{3}$m & 8.4  & 79.0 & 9.17 & 8.6 & 0.072 & 3.9 \\
Mo$_3$Os      & Pm$\bar{3}$n & 14.7 & 59.3 & 9.19 & 6.4 & 0.115 & 3.9 \\
\end{tabular}
\end{ruledtabular}
\end{table}

\begin{table}[b]
\centering
\caption{\label{tab:TopHc2Spin}Superconducting parameters (predicted) for the top-$H_{c2}$ Type-II compounds that include Cr, Mn, V, and Ti.}
\footnotesize
\begin{ruledtabular}
\begin{tabular}{l c c c c c c c c}
Material & Spcgrp. & $\Tc^{\mathrm{El}}$~(K) & $\lambdaL$~(nm) & $\xi$~(nm) & $\kappa$ & $H_{c1}$~(T) & $H_{c2}$~(T) \\
\hline
Cr$_4$NbRe   & F$\bar{4}$3m   & 19.9 & 86.0 & 2.22 & 38.8 & 0.093 & 66.9 \\
Cr$_4$MoTc   & F$\bar{4}$3m   & 19.9 & 75.7 & 3.99 & 19.0 & 0.099 & 20.7 \\
NbTiV$_4$    & F$\bar{4}$3m   & 15.9 & 79.8 & 4.24 & 18.8 & 0.089 & 18.3 \\
Cr$_4$TcW    & F$\bar{4}$3m   & 19.6 & 74.7 & 4.30 & 17.3 & 0.100 & 17.8 \\
Cr$_4$NbOs   & F$\bar{4}$3m   & 18.5 & 78.0 & 4.35 & 18.0 & 0.092 & 17.4 \\
Cr$_3$Ru     & Pm$\bar{3}$n   & 18.9 & 69.8 & 4.36 & 16.0 & 0.111 & 17.3 \\
Cr$_4$ReW    & F$\bar{4}$3m   & 19.4 & 75.6 & 4.49 & 16.8 & 0.096 & 16.3 \\
Cr$_4$OsTa   & F$\bar{4}$3m   & 17.5 & 70.8 & 4.83 & 14.7 & 0.105 & 14.1 \\
Cr$_4$NbW    & P1             & 17.8 & 96.2 & 5.30 & 18.1 & 0.061 & 11.7 \\
Cr$_3$Os     & Pm$\bar{3}$n   & 18.1 & 61.5 & 6.30 &  9.8 & 0.123 &  8.3 \\
Ti$_2$W      & I4/mmm         & 15.3 & 56.7 & 7.10 &  8.0 & 0.135 &  6.5 \\
Nb$_3$Sn$_2$Ti$_3$ & R32      & 13.7 & 96.3 & 7.16 & 13.5 & 0.056 &  6.4 \\
Mn$_4$MoV    & F$\bar{4}$3m   & 16.0 & 63.7 & 7.25 &  8.8 & 0.111 &  6.3 \\
HfTi$_3$H    & Pm$\bar{3}$m   & 17.6 & 52.8 & 7.31 &  7.2 & 0.151 &  6.2 \\
Cr$_4$TaW    & F$\bar{4}$3m   & 17.4 & 68.7 & 7.37 &  9.3 & 0.097 &  6.1 \\
\end{tabular}
\end{ruledtabular}
\end{table}

\subsection{The highest critical field materials and the technologically relevant regime}

Tables \ref{tab:TopHc2NoSpin} and \ref{tab:TopHc2Spin} list the top-$H_{c2}$ Type-II superconductors in our database,
distinguished by the constituent elements.
All of them are found deep in the Type-II regime ($\kappa \gg 1/\sqrt{2}$).
In Table~\ref{tab:TopHc2NoSpin},
the highest $H_{c2}$ is observed for LiMoN$_2$ (48~T), whose short coherence length (2.62~nm) and R3m symmetry distinguish it sharply from the rest of the materials.
A comparison between LiMoN$_2$ and LiTcN$_2$ shows that %
electron doping alters the space group, nearly doubles $\xi$, and thus suppresses $H_{c2}$ by a factor of four;
yet at similar $\kappa$ due to the enhanced $\lambdaL$.
Interestingly, we find several Nb-based compounds in this list that crystallize in cubic, tetragonal, and hexagonal space groups.
In particular, Nb$_3$Pt (Pm$\bar{3}$n) stands out due to a predicted $\Tc \sim 13$~K and a strong $H_{c2} \sim 10$~T.

In Table~\ref{tab:TopHc2Spin}, 
which focuses specifically on materials that include Cr, Mn, V, and Ti,
the upper critical fields reach even higher values.
We observe many Cr-based compounds in this list, which are characterized by low Fermi velocities (see Figs.~\ref{fig:Overview} and~\ref{fig:HcKappa}).
The highest $H_{c2}$ values are found in F$\bar{4}$3m-type compounds, most notably Cr$_4$NbRe, which reaches an exceptionally large $H_{c2} \sim 67$~T.
Within this structural family, substitutions that increase the coherence length systematically suppress $H_{c2}$, as seen when moving from Cr$_4$NbRe ($\xi = 2.22$~nm) to compounds such as Cr$_4$MoTc and Cr$_4$TcW ($\xi \sim 4$~nm) or Cr$_4$TaW ($\xi = 7.37$~nm).

Although phases with non-zero magnetic moments were removed from the $\alpha^2F(\omega)$ databases \cite{betenet, Cerqueira2023},
we speculate that some of these compounds may host subtle antiferromagnetic instabilities. However, Cr$_3$Ru is an experimentally established superconductor with $H_{c2} \sim 5$~T \cite{Cr3Ru:2022}, and the prominence of related materials among the predicted high-critical-field compounds points to a potentially important trend. We therefore list them separately, along with additional systems for which spin fluctuations may play a role. %

Intriguingly, already within the first dataset presented here, we identify several promising candidates
in or near the stringent technologically relevant regime, which we defined above by $\Tc \gtrsim 20$~K and $H_{c2} \gtrsim 10$~T.
Specifically, this comprises many of the Cr-based F$\bar{4}$3m materials listed in Tables \ref{tab:TopHc2Spin}, particularly Cr$_4$NbRe,
but also LiMoN$_2$, LiTcN$_2$,
and particularly several Nb-based compounds from Table~\ref{tab:TopHc2NoSpin} including the well-known superconductor Nb$_3$Pt.
A figure showing $H_{c1}$ and $H_{c2}$ as functions of $\Tc$ is provided in the Supplemental Information.

\begin{figure*}
\begin{center}
\includegraphics[width=\linewidth]{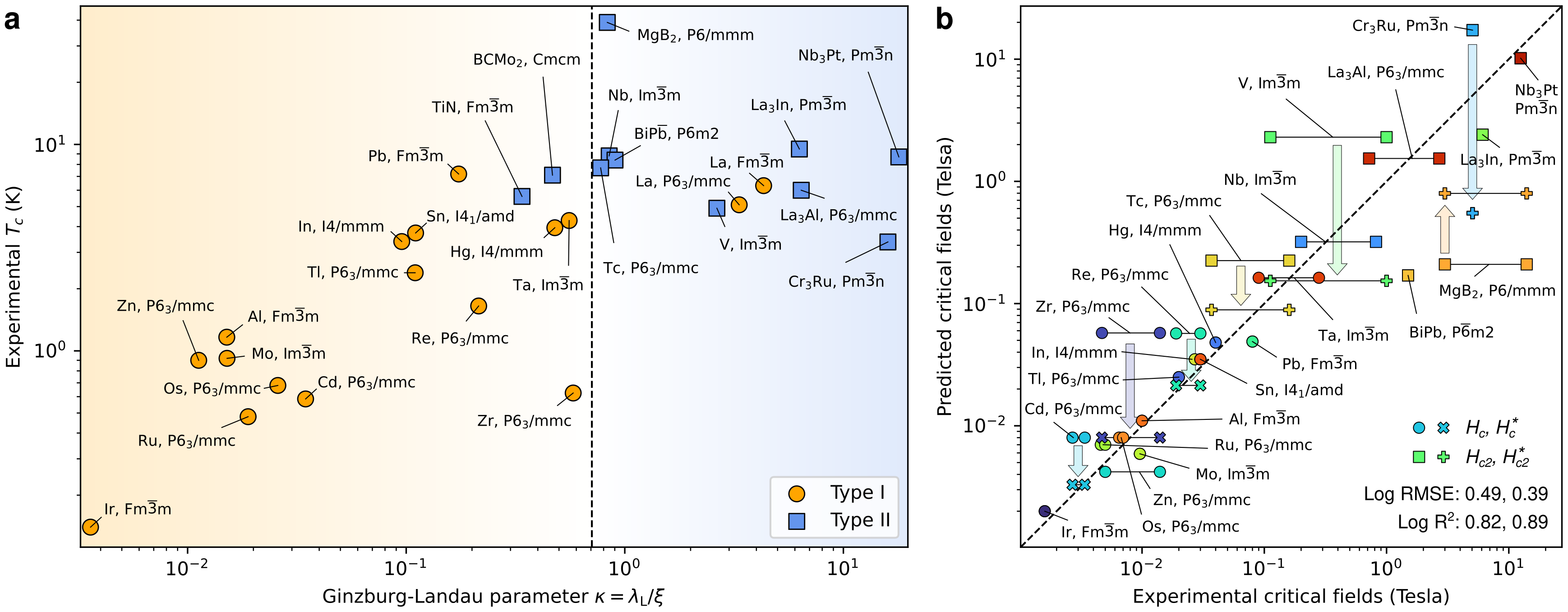}
\caption{\label{fig:ExpComp}
\textbf{\boldmath Comparison of predicted and experimental critical field properties.}
\textbf{a}~Analysis of the Type-I (orange) versus Type-II (blue) superconductor classification, based on the Ginzburg-Landau parameter $\kappa$ predicted entirely from first principles.
The vertical axis corresponds to the experimental $\Tc$, which shows that the classification cannot be trivially deduced from the critical temperature.
\textbf{b}~Parity plot of predicted versus experimental critical fields ($H_{c}$: circles; $H_{c2}$: squares).
A table of the data, compiled from Refs.~\cite{Roberts-SC:1976, Eisenstein:1954, MatthiasGeballeCompton:1963, Schwochau-Tc:2000, MgB2-Hc2:2002, Cr3Ru:2022}, is provided in the Supplemental Information.
For those materials where the predicted $\Tc$ is in significant disagreement with the experimental value, we show in addition a rescaled critical field value using the experimental $\Tc$ ($H_{c}^{*}$: crosses; $H_{c2}^{*}$: pluses), which generally enhances the quantitative agreement as evidenced by the log RMSE and R$^2$ metrics.
}
\end{center}
\end{figure*}

\subsection{Comparison of predicted and experimental critical fields and key role of electron-phonon renormalizations}

Finally, we assess the accuracy of our workflow by comparing the predicted superconducting parameters to experimental measurements.
First, we analyze the Type classification based on the calculated $\kappa$ [Fig.~\ref{fig:ExpComp}(a)].
Intriguingly, the predicted classification agrees with experiment for most materials in our representative test set, %
which comprises elemental superconductors, a number of A15 compounds, BiPb, and the layered superconductor MgB$_2$.
As prominent examples, Pb is correctly predicted as Type~I, and Nb and MgB$_2$ are correctly identified as Type~II.
We find MgB$_2$ close to the threshold, which may be rather conservative due to the underpredicted $\Tc$ (see below).
Considering that all quantities are obtained exclusively from DFT, this accuracy is still remarkable.

Second, we compare predicted versus experimental critical fields $H_{c}$ (Type~I) and $H_{c2}$ (Type~II) quantitatively over four orders of magnitude [Fig.~\ref{fig:ExpComp}(b)]. %
The range given for the experimental values reflects their spread in the literature.
We find that the agreement ranges from very good (Pb, Nb) to within an order of magnitude (MgB$_2$).
We identify a central error source to be $\Tc$, particularly for $H_{c2}$, where any $\Tc$ deviation propagates quadratically.
An example is V, where spin fluctuations are expected to play a substantial role. %
If we rescale the critical fields to the experimental $\Tc$, we achieve an improved agreement for several additional materials. %
Since the electronic properties are predicted rather accurately, even across different exchange-correlation functionals (see Methods),
additional uncertainties remain significantly smaller than those introduced by $\Tc$ itself.

One of our most important findings is that a simple BCS approach is not sufficient:
In addition to strong-coupling corrections to the superconducting gap,
electron-phonon renormalizations of the effective mass are key
to obtaining the correct order of magnitude for the critical fields.
For Nb, we observe an enhancement of the predicted London penetration depth from $\lambdaL = 19$ to $27~\mathrm{nm}$ (experiment: $\sim 39~\mathrm{nm}$ \cite{Kittel:2007}) if considering electron-phonon renormalizations.
At the same time, the predicted coherence length, evaluated by using the Eliashberg gap, decreases from $\xi = 68$ to $32~\mathrm{nm}$ (experiment: $\sim 38~\mathrm{nm}$).
As a result, the Ginzburg-Landau parameter increases from $\kappa = 0.28$ to $0.85$, bringing it much closer to the experimental value of $\sim 1.03$
and changing the theoretical classification of Nb from a \mbox{Type-I} to a Type-II superconductor, in agreement with experiment.
A similar behavior is found for Pb, where $\lambdaL$ increases from $20$ to $29~\mathrm{nm}$ (experiment: $\sim 37~\mathrm{nm}$ \cite{Kittel:2007}),
while $\xi$ is reduced from $339$ to $165~\mathrm{nm}$ (experiment: $\sim 83~\mathrm{nm}$).
Consequently, the Ginzburg-Landau parameter is enhanced from $\kappa = 0.06$ to $0.17$ (experiment: $\sim 0.45$).
In both materials, the calculated $\lambdaL$ remains systematically smaller than the experimental values, as expected.
This leads to an underestimation of $\kappa$, yet significantly less severe than in the unrenormalized case.
Ultimately, the predicted upper critical field $H_{c2}$ of Nb is enhanced from 0.072 to 0.321~T (experiment: 0.2--0.8~T~\cite{Roberts-SC:1976, Eisenstein:1954, MatthiasGeballeCompton:1963}) due to renormalization effects; similarly, the predicted $H_{c}$ of Pb increases from 0.034 to 0.049~T (experiment: 0.08~T).

While the critical-field prediction is convincingly accurate for several Type-I and Type-II superconductors such as Al, Pb, Hg, Nb, Nb$_3$Pt, and also many hcp metals (Fig.~\ref{fig:ExpComp}),
MgB$_2$ emerges as interesting challenge, which highlights both the complexity of its physics from the critical-field perspective as well as opportunities for future methodological innovation. %
The predicted $\Tc$ using the present workflow is half of the experimental value, similar to earlier work~\cite{Xie2019}.
If we correct the upper critical field by using the experimental $\Tc$, %
we obtain an enhancement from $\sim 0.2$ to $\sim 0.8$~T,
which still deviates significantly from the experimental range of $3$~T ($c$~axis) to $14$~T ($ab$~plane) \cite{MgB2-Hc2:2002}.
Possible reasons are the anisotropy and multiband nature of this material \cite{Choi2002, Margine2013}.
In addition to anisotropic Fermi velocity components, the electron-phonon interactions and the resulting effective-mass renormalizations are expected to vary between different directions %
and different sheets of the Fermi surface.
We speculate that accounting for these effects will considerably enhance the predictions for strongly anisotropic materials.

\section{Discussion}

In this work, we have taken a first step towards the discovery of superconductors with high critical magnetic fields by developing a high-throughput workflow that combines clean-limit Eliashberg theory with density functional theory.
By applying this framework to more than $7{,}300$ electron-phonon-paired superconductors across diverse materials classes, we constructed a consistent database of critical fields $H_c$, $H_{c1}$, and $H_{c2}$ together with the underlying electronic and superconducting properties.
This unified perspective reveals previously unrecognized statistical trends in superconducting magnetic response:
We uncovered an unexpectedly large number of Type-I superconductors and showed that larger and more complex crystal structures generically support higher critical fields and Type-II behavior.
Moreover, we pointed out the importance of strong-coupling corrections and electron-phonon renormalizations of the effective mass for predictions of critical fields.
In stark contrast, a simple BCS approach proved to be insufficient.

The resulting dataset comprises accurately sampled Fermi surfaces, averaged Fermi velocities, densities of states, Eliashberg critical temperatures and superconducting gaps, coherence lengths, London penetration depths, Ginzburg-Landau parameters, and critical magnetic fields. As such, it naturally complements existing first-principles databases that provide optimized crystal structures and Eliashberg spectral functions, extending them to magnetic-field-related superconducting properties that have so far remained unexplored at scale.

Beyond its fundamental significance, the present framework has direct technological relevance. The data allow access to the superheating field, a key quantity for minimizing power losses in superconducting RF cavities, and highlight materials with high critical fields that are of interest for compact and efficient electron-beam lithography systems in the semiconductor industry
and for high-performance superconducting magnets for high-field laboratories, fusion reactors, particle accelerators, and MRI machines.
More broadly, the insights obtained here are relevant for identifying high-performance superconductors with small unit cells and three-dimensional electronic structure.

Finally, by establishing a quantitative link between electronic structure, pairing, and magnetic response across a large materials space, this work lays the foundation for future AI-driven approaches to inverse materials design, enabling the targeted discovery of superconductors with simultaneously enhanced critical temperatures and critical magnetic fields.

\section{Methods}

For each of the $\sim 7{,}300$ compounds in the database,
we performed an independent DFT structural optimization
using the \textsc{VASP} code \cite{Kresse1996, Kresse1996b, Kresse1999} in conjunction with the PBEsol exchange-correlation functional \cite{Perdew2008}
to confirm earlier findings,
followed by a self-consistent and a non-self-consistent calculation
to obtain accurate Fermi surfaces and densities of states.
We use dense $k$-point grids that are scaled with the size of the unit cell (i.e., keeping a fixed density of $k$~points).
For example, we employ $40\times40\times35$ for MgB$_2$ and $45\times45\times45$ for Nb,
which are further interpolated on a five times denser $k$-point grid before Fermi surface plotting and Fermi velocity evaluation
using a modified version of the \textsc{ifermi} code.
We found this density to balance accuracy and feasibility in our high-throughput strategy.
Materials from the earlier databases that we identified here to exhibit a band gap have been removed.

From extensive convergence tests for a subset of over $6{,}400$ compounds from our database,
we conclude that the $k$-grid convergence achieved by our methodology is statistically (i.e., considering the standard deviation) within
$\sim \pm 2\%$ for $n(E_\text{F})$ and
$\sim \pm 1\%$ for $v_\text{F}$,
which is considerably more accurate than typical $\Tc$ predictions.
Moreover, for a test set of 17 elemental metals,
we reproduce earlier $v_\text{F}$ DFT predictions \cite{Gall-FermiVelocitiesMetals:16} within $\sim \pm 1\%$.
Moreover, we found that the statistical variations of the predicted $v_\text{F}$ between the exchange-correlation functionals LDA, PBE, and PBEsol are within $\sim \pm 3\%$.

We used the EPW code \cite{EPW-1, EPW-2},
embedded in a high-throughput framework we developed in \textsc{Python},
to solve the isotropic Eliashberg equations~\cite{EliashbergInteractionBetweenElAndLatticeVibrInASC1960, Carbotte1990RMP} for the $\sim 7{,}300$ electron-phonon superconductors,
which are characterized by their Eliashberg spectral function $\alpha^2F(\omega)$:
\begin{align}
Z(i\omega_n)
  &= 1 + \frac{\pi T}{\omega_n}
     \sum_{m=-\infty}^{\infty}
     \frac{\lambda_{n,m} \ \omega_m}{\sqrt{\omega_m^2 + \Delta^2(i\omega_m)}},
\\[6pt]
Z(i\omega_n) \Delta(i\omega_n)
   &= \pi T \sum_{m=-\infty}^{\infty} \cdots \\
   &\cdots \frac{[ \lambda_{n,m} - \mu^{*}_{\text{El}} \, \Theta(\omega_\text{c} - \vert\omega_m\vert)] \, \Delta(i\omega_m)}{\sqrt{\omega_m^2 + \Delta^2(i\omega_m)}},
\end{align}
where
\begin{align}
\lambda_{n,m}
  &= 2 \int_0^{\infty} \text{d}\omega \,
    \frac{\omega \ \alpha^2 F(\omega)}
         {\omega^2 + (\omega_n - \omega_m)^2 },
\\[6pt]
i\omega_n &= i(2n+1)\pi T \quad (n \in \mathbb{Z}).
\end{align}
Here we determine the Coulomb parameter $\mu^{*}_{\text{El}}$ dynamically for each material
and rescale it appropriately using a strategy motivated by Pellegrini and Sanna \cite{PellegriniSanna:2024} and earlier considerations by Allen and Dynes \cite{Allen-Dynes1975}:
\begin{equation}
\frac{1}{\mu^{*}_{\text{El}}}
= \frac{1}{\mu^{*}_{\text{AD}}}
+ \ln\!\left( \frac{\omega_{\text{ph}}}{\omega_\text{c}} \right),
\end{equation}
in which $\omega_{\text{ph}}$ is the maximum frequency of the phonon spectrum
and $\omega_\text{c} = 10 \, \omega_2$ is the cutoff for the Matsubara frequencies
used during solving the Eliashberg equations.
In this expression, we set $\mu^{*}_{\text{AD}} = 0.14$ for materials containing transition metals and $\mu^{*}_{\text{AD}} = 0.10$ otherwise.

For materials where solving the Eliashberg equations did not converge,
usually in the case of very low $\Tc \ll 1$~K,
we resort to Allen-Dynes $\Tc$ and/or BCS gaps,
following the philosophy of predicting critical magnetic fields for as many materials as possible.

While there are individually tailored strategies for some compounds that render a $\Tc$ closer to experiment,
e.g., by solving the anisotropic Eliashberg equations \cite{Choi2002, Margine2013},
their applicability in a high-throughput approach is so far unclear.
Here, we report the high-throughput values obtained consistently for all materials.

\section{Acknowledgments}

We thank A.~Gurevich, D.~Larbalestier, and R.~Margine for helpful discussions.
B.G., J.J.H., G.R.S., R.G.H., and P.J.H.\ acknowledge funding by Grant No.~NSF-DMREF-2522891.
B.G., R.G.H., and P.J.H.\ acknowledge computational resources provided by the University of Florida via the 'AI and Complex Computational Research' program.
P.M.D.~acknowledges support from the Laboratory Directed Research and Development Program of Oak Ridge National Laboratory, managed by UT-Battelle, LLC, for the US Department of Energy. Notice of Copyright: This manuscript has been authored by UT-Battelle, LLC, under contract DE-AC05-00OR22725 with the US Department of Energy (DOE). The publisher acknowledges the US government license to provide public access under the DOE Public Access Plan (http://energy.gov/downloads/doe-public-access-plan).

\section{Author Contributions}

B.G., P.M.D., R.G.H., and P.J.H.\ conceived the project and conceptualized the framework.
B.G.\ performed the DFT and Eliashberg calculations, implemented the high-thoughtput workflows, constructed the database, and performed the data analysis and visualization.
J.J.H.\ and G.R.S.\ contributed experimental insights on the predictions and helped compiling the experimental reference data.
B.G.\ and P.J.H.\ wrote the manuscript.
All authors discussed the results and contributed to revising and editing the manuscript.

\section*{Competing Interests}

The authors declare no competing interests.

\section{Data Availability}

The critical field database is available at \url{https://github.com/henniggroup}.


\begin{thebibliography}{58}%
\makeatletter
\providecommand \@ifxundefined [1]{%
 \@ifx{#1\undefined}
}%
\providecommand \@ifnum [1]{%
 \ifnum #1\expandafter \@firstoftwo
 \else \expandafter \@secondoftwo
 \fi
}%
\providecommand \@ifx [1]{%
 \ifx #1\expandafter \@firstoftwo
 \else \expandafter \@secondoftwo
 \fi
}%
\providecommand \natexlab [1]{#1}%
\providecommand \enquote  [1]{``#1''}%
\providecommand \bibnamefont  [1]{#1}%
\providecommand \bibfnamefont [1]{#1}%
\providecommand \citenamefont [1]{#1}%
\providecommand \href@noop [0]{\@secondoftwo}%
\providecommand \href [0]{\begingroup \@sanitize@url \@href}%
\providecommand \@href[1]{\@@startlink{#1}\@@href}%
\providecommand \@@href[1]{\endgroup#1\@@endlink}%
\providecommand \@sanitize@url [0]{\catcode `\\12\catcode `\$12\catcode `\&12\catcode `\#12\catcode `\^12\catcode `\_12\catcode `\%12\relax}%
\providecommand \@@startlink[1]{}%
\providecommand \@@endlink[0]{}%
\providecommand \url  [0]{\begingroup\@sanitize@url \@url }%
\providecommand \@url [1]{\endgroup\@href {#1}{\urlprefix }}%
\providecommand \urlprefix  [0]{URL }%
\providecommand \Eprint [0]{\href }%
\providecommand \doibase [0]{https://doi.org/}%
\providecommand \selectlanguage [0]{\@gobble}%
\providecommand \bibinfo  [0]{\@secondoftwo}%
\providecommand \bibfield  [0]{\@secondoftwo}%
\providecommand \translation [1]{[#1]}%
\providecommand \BibitemOpen [0]{}%
\providecommand \bibitemStop [0]{}%
\providecommand \bibitemNoStop [0]{.\EOS\space}%
\providecommand \EOS [0]{\spacefactor3000\relax}%
\providecommand \BibitemShut  [1]{\csname bibitem#1\endcsname}%
\let\auto@bib@innerbib\@empty
\bibitem [{\citenamefont {Bednorz}\ and\ \citenamefont {M{\"u}ller}(1986)}]{Bednorz1986}%
  \BibitemOpen
  \bibfield  {author} {\bibinfo {author} {\bibfnamefont {J.~G.}\ \bibnamefont {Bednorz}}\ and\ \bibinfo {author} {\bibfnamefont {K.~A.}\ \bibnamefont {M{\"u}ller}},\ }\bibfield  {title} {\bibinfo {title} {{Possible high-$T_{c}$ superconductivity in the Ba-La-Cu-O system}},\ }\href {https://doi.org/10.1007/BF01303701} {\bibfield  {journal} {\bibinfo  {journal} {Zeitschrift f{\"u}r Physik B Condensed Matter}\ }\textbf {\bibinfo {volume} {64}},\ \bibinfo {pages} {189} (\bibinfo {year} {1986})}\BibitemShut {NoStop}%
\bibitem [{\citenamefont {Wu}\ \emph {et~al.}(1987)\citenamefont {Wu}, \citenamefont {Ashburn}, \citenamefont {Torng}, \citenamefont {Hor}, \citenamefont {Meng}, \citenamefont {Gao}, \citenamefont {Huang}, \citenamefont {Wang},\ and\ \citenamefont {Chu}}]{Wu1987}%
  \BibitemOpen
  \bibfield  {author} {\bibinfo {author} {\bibfnamefont {M.~K.}\ \bibnamefont {Wu}}, \bibinfo {author} {\bibfnamefont {J.~R.}\ \bibnamefont {Ashburn}}, \bibinfo {author} {\bibfnamefont {C.~J.}\ \bibnamefont {Torng}}, \bibinfo {author} {\bibfnamefont {P.~H.}\ \bibnamefont {Hor}}, \bibinfo {author} {\bibfnamefont {R.~L.}\ \bibnamefont {Meng}}, \bibinfo {author} {\bibfnamefont {L.}~\bibnamefont {Gao}}, \bibinfo {author} {\bibfnamefont {Z.~J.}\ \bibnamefont {Huang}}, \bibinfo {author} {\bibfnamefont {Y.~Q.}\ \bibnamefont {Wang}},\ and\ \bibinfo {author} {\bibfnamefont {C.~W.}\ \bibnamefont {Chu}},\ }\bibfield  {title} {\bibinfo {title} {{Superconductivity at 93 K in a new mixed-phase Y-Ba-Cu-O compound system at ambient pressure}},\ }\href {https://doi.org/10.1103/PhysRevLett.58.908} {\bibfield  {journal} {\bibinfo  {journal} {Phys. Rev. Lett.}\ }\textbf {\bibinfo {volume} {58}},\ \bibinfo {pages} {908} (\bibinfo {year} {1987})}\BibitemShut {NoStop}%
\bibitem [{\citenamefont {Schilling}\ \emph {et~al.}(1993)\citenamefont {Schilling}, \citenamefont {Cantoni}, \citenamefont {Guo},\ and\ \citenamefont {Ott}}]{Schilling1993}%
  \BibitemOpen
  \bibfield  {author} {\bibinfo {author} {\bibfnamefont {A.}~\bibnamefont {Schilling}}, \bibinfo {author} {\bibfnamefont {M.}~\bibnamefont {Cantoni}}, \bibinfo {author} {\bibfnamefont {J.~D.}\ \bibnamefont {Guo}},\ and\ \bibinfo {author} {\bibfnamefont {H.~R.}\ \bibnamefont {Ott}},\ }\bibfield  {title} {\bibinfo {title} {{Superconductivity above 130 K in the Hg-Ba-Ca-Cu-O system}},\ }\href {https://doi.org/10.1038/363056a0} {\bibfield  {journal} {\bibinfo  {journal} {Nature}\ }\textbf {\bibinfo {volume} {363}},\ \bibinfo {pages} {56} (\bibinfo {year} {1993})}\BibitemShut {NoStop}%
\bibitem [{\citenamefont {Nagamatsu}\ \emph {et~al.}(2001)\citenamefont {Nagamatsu}, \citenamefont {Nakagawa}, \citenamefont {Muranaka}, \citenamefont {Zenitani},\ and\ \citenamefont {Akimitsu}}]{Nagamatsu2001}%
  \BibitemOpen
  \bibfield  {author} {\bibinfo {author} {\bibfnamefont {J.}~\bibnamefont {Nagamatsu}}, \bibinfo {author} {\bibfnamefont {N.}~\bibnamefont {Nakagawa}}, \bibinfo {author} {\bibfnamefont {T.}~\bibnamefont {Muranaka}}, \bibinfo {author} {\bibfnamefont {Y.}~\bibnamefont {Zenitani}},\ and\ \bibinfo {author} {\bibfnamefont {J.}~\bibnamefont {Akimitsu}},\ }\bibfield  {title} {\bibinfo {title} {Superconductivity at 39{\thinspace}k in magnesium diboride},\ }\href {https://doi.org/10.1038/35065039} {\bibfield  {journal} {\bibinfo  {journal} {Nature}\ }\textbf {\bibinfo {volume} {410}},\ \bibinfo {pages} {63} (\bibinfo {year} {2001})}\BibitemShut {NoStop}%
\bibitem [{\citenamefont {Kamihara}\ \emph {et~al.}(2008)\citenamefont {Kamihara}, \citenamefont {Watanabe}, \citenamefont {Hirano},\ and\ \citenamefont {Hosono}}]{Kamihara2008}%
  \BibitemOpen
  \bibfield  {author} {\bibinfo {author} {\bibfnamefont {Y.}~\bibnamefont {Kamihara}}, \bibinfo {author} {\bibfnamefont {T.}~\bibnamefont {Watanabe}}, \bibinfo {author} {\bibfnamefont {M.}~\bibnamefont {Hirano}},\ and\ \bibinfo {author} {\bibfnamefont {H.}~\bibnamefont {Hosono}},\ }\bibfield  {title} {\bibinfo {title} {{Iron-Based Layered Superconductor La[O$_{1-x}$F$_{x}$]FeAs ($x = 0.05$-$0.12$) with $T_{c} = 26$ K}},\ }\href {https://doi.org/10.1021/ja800073m} {\bibfield  {journal} {\bibinfo  {journal} {Journal of the American Chemical Society}\ }\textbf {\bibinfo {volume} {130}},\ \bibinfo {pages} {3296} (\bibinfo {year} {2008})}\BibitemShut {NoStop}%
\bibitem [{\citenamefont {Stewart}(2011)}]{StewartRMP2011}%
  \BibitemOpen
  \bibfield  {author} {\bibinfo {author} {\bibfnamefont {G.~R.}\ \bibnamefont {Stewart}},\ }\bibfield  {title} {\bibinfo {title} {Superconductivity in iron compounds},\ }\href {https://doi.org/10.1103/RevModPhys.83.1589} {\bibfield  {journal} {\bibinfo  {journal} {Rev. Mod. Phys.}\ }\textbf {\bibinfo {volume} {83}},\ \bibinfo {pages} {1589} (\bibinfo {year} {2011})}\BibitemShut {NoStop}%
\bibitem [{\citenamefont {Cao}\ \emph {et~al.}(2018)\citenamefont {Cao}, \citenamefont {Fatemi}, \citenamefont {Fang}, \citenamefont {Watanabe}, \citenamefont {Taniguchi}, \citenamefont {Kaxiras},\ and\ \citenamefont {Jarillo-Herrero}}]{Cao2018}%
  \BibitemOpen
  \bibfield  {author} {\bibinfo {author} {\bibfnamefont {Y.}~\bibnamefont {Cao}}, \bibinfo {author} {\bibfnamefont {V.}~\bibnamefont {Fatemi}}, \bibinfo {author} {\bibfnamefont {S.}~\bibnamefont {Fang}}, \bibinfo {author} {\bibfnamefont {K.}~\bibnamefont {Watanabe}}, \bibinfo {author} {\bibfnamefont {T.}~\bibnamefont {Taniguchi}}, \bibinfo {author} {\bibfnamefont {E.}~\bibnamefont {Kaxiras}},\ and\ \bibinfo {author} {\bibfnamefont {P.}~\bibnamefont {Jarillo-Herrero}},\ }\bibfield  {title} {\bibinfo {title} {Unconventional superconductivity in magic-angle graphene superlattices},\ }\href {https://doi.org/10.1038/nature26160} {\bibfield  {journal} {\bibinfo  {journal} {Nature}\ }\textbf {\bibinfo {volume} {556}},\ \bibinfo {pages} {43} (\bibinfo {year} {2018})}\BibitemShut {NoStop}%
\bibitem [{\citenamefont {Kamysbayev}\ \emph {et~al.}(2020)\citenamefont {Kamysbayev}, \citenamefont {Filatov}, \citenamefont {Hu}, \citenamefont {Rui}, \citenamefont {Lagunas}, \citenamefont {Wang}, \citenamefont {Klie},\ and\ \citenamefont {Talapin}}]{Kamysbayev2020}%
  \BibitemOpen
  \bibfield  {author} {\bibinfo {author} {\bibfnamefont {V.}~\bibnamefont {Kamysbayev}}, \bibinfo {author} {\bibfnamefont {A.~S.}\ \bibnamefont {Filatov}}, \bibinfo {author} {\bibfnamefont {H.}~\bibnamefont {Hu}}, \bibinfo {author} {\bibfnamefont {X.}~\bibnamefont {Rui}}, \bibinfo {author} {\bibfnamefont {F.}~\bibnamefont {Lagunas}}, \bibinfo {author} {\bibfnamefont {D.}~\bibnamefont {Wang}}, \bibinfo {author} {\bibfnamefont {R.~F.}\ \bibnamefont {Klie}},\ and\ \bibinfo {author} {\bibfnamefont {D.~V.}\ \bibnamefont {Talapin}},\ }\bibfield  {title} {\bibinfo {title} {{Covalent surface modifications and superconductivity of two-dimensional metal carbide MXenes}},\ }\href {https://doi.org/10.1126/science.aba8311} {\bibfield  {journal} {\bibinfo  {journal} {Science}\ }\textbf {\bibinfo {volume} {369}},\ \bibinfo {pages} {979} (\bibinfo {year} {2020})}\BibitemShut {NoStop}%
\bibitem [{\citenamefont {Flores-Livas}\ \emph {et~al.}(2020)\citenamefont {Flores-Livas}, \citenamefont {Boeri}, \citenamefont {Sanna}, \citenamefont {Profeta}, \citenamefont {Arita},\ and\ \citenamefont {Eremets}}]{FLORESLIVAS2020}%
  \BibitemOpen
  \bibfield  {author} {\bibinfo {author} {\bibfnamefont {J.~A.}\ \bibnamefont {Flores-Livas}}, \bibinfo {author} {\bibfnamefont {L.}~\bibnamefont {Boeri}}, \bibinfo {author} {\bibfnamefont {A.}~\bibnamefont {Sanna}}, \bibinfo {author} {\bibfnamefont {G.}~\bibnamefont {Profeta}}, \bibinfo {author} {\bibfnamefont {R.}~\bibnamefont {Arita}},\ and\ \bibinfo {author} {\bibfnamefont {M.}~\bibnamefont {Eremets}},\ }\bibfield  {title} {\bibinfo {title} {A perspective on conventional high-temperature superconductors at high pressure: Methods and materials},\ }\href {https://doi.org/https://doi.org/10.1016/j.physrep.2020.02.003} {\bibfield  {journal} {\bibinfo  {journal} {Physics Reports}\ }\textbf {\bibinfo {volume} {856}},\ \bibinfo {pages} {1} (\bibinfo {year} {2020})}\BibitemShut {NoStop}%
\bibitem [{\citenamefont {Boeri}\ \emph {et~al.}(2022)\citenamefont {Boeri}, \citenamefont {Hennig}, \citenamefont {Hirschfeld}, \citenamefont {Profeta}, \citenamefont {Sanna}, \citenamefont {Zurek}, \citenamefont {Pickett}, \citenamefont {Amsler}, \citenamefont {Dias}, \citenamefont {Eremets}, \citenamefont {Heil}, \citenamefont {Hemley}, \citenamefont {Liu}, \citenamefont {Ma}, \citenamefont {Pierleoni}, \citenamefont {Kolmogorov}, \citenamefont {Rybin}, \citenamefont {Novoselov}, \citenamefont {Anisimov}, \citenamefont {Oganov}, \citenamefont {Pickard}, \citenamefont {Bi}, \citenamefont {Arita}, \citenamefont {Errea}, \citenamefont {Pellegrini}, \citenamefont {Requist}, \citenamefont {Gross}, \citenamefont {Margine}, \citenamefont {Xie}, \citenamefont {Quan}, \citenamefont {Hire}, \citenamefont {Fanfarillo}, \citenamefont {Stewart}, \citenamefont {Hamlin}, \citenamefont {Stanev}, \citenamefont {Gonnelli}, \citenamefont {Piatti}, \citenamefont {Romanin}, \citenamefont {Daghero},\ and\ \citenamefont
  {Valenti}}]{ROADMAP}%
  \BibitemOpen
  \bibfield  {author} {\bibinfo {author} {\bibfnamefont {L.}~\bibnamefont {Boeri}}, \bibinfo {author} {\bibfnamefont {R.}~\bibnamefont {Hennig}}, \bibinfo {author} {\bibfnamefont {P.}~\bibnamefont {Hirschfeld}}, \bibinfo {author} {\bibfnamefont {G.}~\bibnamefont {Profeta}}, \bibinfo {author} {\bibfnamefont {A.}~\bibnamefont {Sanna}}, \bibinfo {author} {\bibfnamefont {E.}~\bibnamefont {Zurek}}, \bibinfo {author} {\bibfnamefont {W.~E.}\ \bibnamefont {Pickett}}, \bibinfo {author} {\bibfnamefont {M.}~\bibnamefont {Amsler}}, \bibinfo {author} {\bibfnamefont {R.}~\bibnamefont {Dias}}, \bibinfo {author} {\bibfnamefont {M.~I.}\ \bibnamefont {Eremets}}, \bibinfo {author} {\bibfnamefont {C.}~\bibnamefont {Heil}}, \bibinfo {author} {\bibfnamefont {R.~J.}\ \bibnamefont {Hemley}}, \bibinfo {author} {\bibfnamefont {H.}~\bibnamefont {Liu}}, \bibinfo {author} {\bibfnamefont {Y.}~\bibnamefont {Ma}}, \bibinfo {author} {\bibfnamefont {C.}~\bibnamefont {Pierleoni}}, \bibinfo {author} {\bibfnamefont {A.~N.}\ \bibnamefont
  {Kolmogorov}}, \bibinfo {author} {\bibfnamefont {N.}~\bibnamefont {Rybin}}, \bibinfo {author} {\bibfnamefont {D.}~\bibnamefont {Novoselov}}, \bibinfo {author} {\bibfnamefont {V.}~\bibnamefont {Anisimov}}, \bibinfo {author} {\bibfnamefont {A.~R.}\ \bibnamefont {Oganov}}, \bibinfo {author} {\bibfnamefont {C.~J.}\ \bibnamefont {Pickard}}, \bibinfo {author} {\bibfnamefont {T.}~\bibnamefont {Bi}}, \bibinfo {author} {\bibfnamefont {R.}~\bibnamefont {Arita}}, \bibinfo {author} {\bibfnamefont {I.}~\bibnamefont {Errea}}, \bibinfo {author} {\bibfnamefont {C.}~\bibnamefont {Pellegrini}}, \bibinfo {author} {\bibfnamefont {R.}~\bibnamefont {Requist}}, \bibinfo {author} {\bibfnamefont {E.~K.~U.}\ \bibnamefont {Gross}}, \bibinfo {author} {\bibfnamefont {E.~R.}\ \bibnamefont {Margine}}, \bibinfo {author} {\bibfnamefont {S.~R.}\ \bibnamefont {Xie}}, \bibinfo {author} {\bibfnamefont {Y.}~\bibnamefont {Quan}}, \bibinfo {author} {\bibfnamefont {A.}~\bibnamefont {Hire}}, \bibinfo {author} {\bibfnamefont {L.}~\bibnamefont
  {Fanfarillo}}, \bibinfo {author} {\bibfnamefont {G.~R.}\ \bibnamefont {Stewart}}, \bibinfo {author} {\bibfnamefont {J.~J.}\ \bibnamefont {Hamlin}}, \bibinfo {author} {\bibfnamefont {V.}~\bibnamefont {Stanev}}, \bibinfo {author} {\bibfnamefont {R.~S.}\ \bibnamefont {Gonnelli}}, \bibinfo {author} {\bibfnamefont {E.}~\bibnamefont {Piatti}}, \bibinfo {author} {\bibfnamefont {D.}~\bibnamefont {Romanin}}, \bibinfo {author} {\bibfnamefont {D.}~\bibnamefont {Daghero}},\ and\ \bibinfo {author} {\bibfnamefont {R.}~\bibnamefont {Valenti}},\ }\bibfield  {title} {\bibinfo {title} {The 2021 room-temperature superconductivity roadmap},\ }\href {https://doi.org/10.1088/1361-648X/ac2864} {\bibfield  {journal} {\bibinfo  {journal} {Journal of Physics: Condensed Matter}\ }\textbf {\bibinfo {volume} {34}},\ \bibinfo {pages} {183002} (\bibinfo {year} {2022})}\BibitemShut {NoStop}%
\bibitem [{\citenamefont {Molodyk}\ and\ \citenamefont {Larbalestier}(2023)}]{Molodyk2023}%
  \BibitemOpen
  \bibfield  {author} {\bibinfo {author} {\bibfnamefont {A.}~\bibnamefont {Molodyk}}\ and\ \bibinfo {author} {\bibfnamefont {D.~C.}\ \bibnamefont {Larbalestier}},\ }\bibfield  {title} {\bibinfo {title} {The prospects of high-temperature superconductors},\ }\href {https://doi.org/10.1126/science.abq4137} {\bibfield  {journal} {\bibinfo  {journal} {Science}\ }\textbf {\bibinfo {volume} {380}},\ \bibinfo {pages} {1220} (\bibinfo {year} {2023})}\BibitemShut {NoStop}%
\bibitem [{\citenamefont {Pellegrini}\ and\ \citenamefont {Sanna}(2024)}]{PellegriniSanna:2024}%
  \BibitemOpen
  \bibfield  {author} {\bibinfo {author} {\bibfnamefont {C.}~\bibnamefont {Pellegrini}}\ and\ \bibinfo {author} {\bibfnamefont {A.}~\bibnamefont {Sanna}},\ }\bibfield  {title} {\bibinfo {title} {Ab initio methods for superconductivity},\ }\href {https://doi.org/10.1038/s42254-024-00738-9} {\bibfield  {journal} {\bibinfo  {journal} {Nature Reviews Physics}\ }\textbf {\bibinfo {volume} {6}},\ \bibinfo {pages} {509} (\bibinfo {year} {2024})}\BibitemShut {NoStop}%
\bibitem [{\citenamefont {Gao}\ \emph {et~al.}(2025)\citenamefont {Gao}, \citenamefont {Cerqueira}, \citenamefont {Sanna}, \citenamefont {Fang}, \citenamefont {Dangi{\'{c}}}, \citenamefont {Errea}, \citenamefont {Wang}, \citenamefont {Botti},\ and\ \citenamefont {Marques}}]{Gao2025}%
  \BibitemOpen
  \bibfield  {author} {\bibinfo {author} {\bibfnamefont {K.}~\bibnamefont {Gao}}, \bibinfo {author} {\bibfnamefont {T.~F.~T.}\ \bibnamefont {Cerqueira}}, \bibinfo {author} {\bibfnamefont {A.}~\bibnamefont {Sanna}}, \bibinfo {author} {\bibfnamefont {Y.-W.}\ \bibnamefont {Fang}}, \bibinfo {author} {\bibfnamefont {{\DJ}.}~\bibnamefont {Dangi{\'{c}}}}, \bibinfo {author} {\bibfnamefont {I.}~\bibnamefont {Errea}}, \bibinfo {author} {\bibfnamefont {H.-C.}\ \bibnamefont {Wang}}, \bibinfo {author} {\bibfnamefont {S.}~\bibnamefont {Botti}},\ and\ \bibinfo {author} {\bibfnamefont {M.~A.~L.}\ \bibnamefont {Marques}},\ }\bibfield  {title} {\bibinfo {title} {The maximum {Tc} of conventional superconductors at ambient pressure},\ }\href {https://doi.org/10.1038/s41467-025-63702-w} {\bibfield  {journal} {\bibinfo  {journal} {Nature Communications}\ }\textbf {\bibinfo {volume} {16}},\ \bibinfo {pages} {8253} (\bibinfo {year} {2025})}\BibitemShut {NoStop}%
\bibitem [{\citenamefont {{Duan}}\ \emph {et~al.}(2014)\citenamefont {{Duan}}, \citenamefont {{Liu}}, \citenamefont {{Tian}}, \citenamefont {{Li}}, \citenamefont {{Huang}}, \citenamefont {{Zhao}}, \citenamefont {{Yu}}, \citenamefont {{Liu}}, \citenamefont {{Tian}},\ and\ \citenamefont {{Cui}}}]{Duan14}%
  \BibitemOpen
  \bibfield  {author} {\bibinfo {author} {\bibfnamefont {D.}~\bibnamefont {{Duan}}}, \bibinfo {author} {\bibfnamefont {Y.}~\bibnamefont {{Liu}}}, \bibinfo {author} {\bibfnamefont {F.}~\bibnamefont {{Tian}}}, \bibinfo {author} {\bibfnamefont {D.}~\bibnamefont {{Li}}}, \bibinfo {author} {\bibfnamefont {X.}~\bibnamefont {{Huang}}}, \bibinfo {author} {\bibfnamefont {Z.}~\bibnamefont {{Zhao}}}, \bibinfo {author} {\bibfnamefont {H.}~\bibnamefont {{Yu}}}, \bibinfo {author} {\bibfnamefont {B.}~\bibnamefont {{Liu}}}, \bibinfo {author} {\bibfnamefont {W.}~\bibnamefont {{Tian}}},\ and\ \bibinfo {author} {\bibfnamefont {T.}~\bibnamefont {{Cui}}},\ }\bibfield  {title} {\bibinfo {title} {{Pressure-induced metallization of dense (H$_{2}$S)$_{2}$H$_{2}$ with high-T$_{c}$ superconductivity}},\ }\href {https://doi.org/10.1038/srep06968} {\bibfield  {journal} {\bibinfo  {journal} {Sci. Rep.}\ }\textbf {\bibinfo {volume} {4}},\ \bibinfo {eid} {6968} (\bibinfo {year} {2014})}\BibitemShut {NoStop}%
\bibitem [{\citenamefont {Drozdov}\ \emph {et~al.}(2015)\citenamefont {Drozdov}, \citenamefont {Eremets}, \citenamefont {Troyan}, \citenamefont {Ksenofontov},\ and\ \citenamefont {Shylin}}]{Drozdov_H3S_2015}%
  \BibitemOpen
  \bibfield  {author} {\bibinfo {author} {\bibfnamefont {A.~P.}\ \bibnamefont {Drozdov}}, \bibinfo {author} {\bibfnamefont {M.~I.}\ \bibnamefont {Eremets}}, \bibinfo {author} {\bibfnamefont {I.~A.}\ \bibnamefont {Troyan}}, \bibinfo {author} {\bibfnamefont {V.}~\bibnamefont {Ksenofontov}},\ and\ \bibinfo {author} {\bibfnamefont {S.~I.}\ \bibnamefont {Shylin}},\ }\bibfield  {title} {\bibinfo {title} {Conventional superconductivity at 203 kelvin at high pressures in the sulfur hydride system},\ }\href {https://doi.org/10.1038/nature14964} {\bibfield  {journal} {\bibinfo  {journal} {Nature}\ }\textbf {\bibinfo {volume} {525}},\ \bibinfo {pages} {73} (\bibinfo {year} {2015})}\BibitemShut {NoStop}%
\bibitem [{\citenamefont {Somayazulu}\ \emph {et~al.}(2019)\citenamefont {Somayazulu}, \citenamefont {Ahart}, \citenamefont {Mishra}, \citenamefont {Geballe}, \citenamefont {Baldini}, \citenamefont {Meng}, \citenamefont {Struzhkin},\ and\ \citenamefont {Hemley}}]{HemleyLaH10}%
  \BibitemOpen
  \bibfield  {author} {\bibinfo {author} {\bibfnamefont {M.}~\bibnamefont {Somayazulu}}, \bibinfo {author} {\bibfnamefont {M.}~\bibnamefont {Ahart}}, \bibinfo {author} {\bibfnamefont {A.~K.}\ \bibnamefont {Mishra}}, \bibinfo {author} {\bibfnamefont {Z.~M.}\ \bibnamefont {Geballe}}, \bibinfo {author} {\bibfnamefont {M.}~\bibnamefont {Baldini}}, \bibinfo {author} {\bibfnamefont {Y.}~\bibnamefont {Meng}}, \bibinfo {author} {\bibfnamefont {V.~V.}\ \bibnamefont {Struzhkin}},\ and\ \bibinfo {author} {\bibfnamefont {R.~J.}\ \bibnamefont {Hemley}},\ }\bibfield  {title} {\bibinfo {title} {Evidence for superconductivity above 260 {K} in lanthanum superhydride at megabar pressures},\ }\href {https://doi.org/10.1103/PhysRevLett.122.027001} {\bibfield  {journal} {\bibinfo  {journal} {Phys. Rev. Lett.}\ }\textbf {\bibinfo {volume} {122}},\ \bibinfo {pages} {027001} (\bibinfo {year} {2019})}\BibitemShut {NoStop}%
\bibitem [{\citenamefont {{Flores-Livas}}\ \emph {et~al.}(2019)\citenamefont {{Flores-Livas}}, \citenamefont {{Boeri}}, \citenamefont {{Sanna}}, \citenamefont {{Profeta}}, \citenamefont {{Arita}},\ and\ \citenamefont {{Eremets}}}]{Eremets_hydride_review2019}%
  \BibitemOpen
  \bibfield  {author} {\bibinfo {author} {\bibfnamefont {J.~A.}\ \bibnamefont {{Flores-Livas}}}, \bibinfo {author} {\bibfnamefont {L.}~\bibnamefont {{Boeri}}}, \bibinfo {author} {\bibfnamefont {A.}~\bibnamefont {{Sanna}}}, \bibinfo {author} {\bibfnamefont {G.}~\bibnamefont {{Profeta}}}, \bibinfo {author} {\bibfnamefont {R.}~\bibnamefont {{Arita}}},\ and\ \bibinfo {author} {\bibfnamefont {M.}~\bibnamefont {{Eremets}}},\ }\bibfield  {title} {\bibinfo {title} {{A Perspective on Conventional High-Temperature Superconductors at High Pressure: Methods and Materials}},\ }\href@noop {} {\bibfield  {journal} {\bibinfo  {journal} {arXiv:1905.06693}\ } (\bibinfo {year} {2019})}\BibitemShut {NoStop}%
\bibitem [{\citenamefont {Pickett}(2023)}]{PickettRMP}%
  \BibitemOpen
  \bibfield  {author} {\bibinfo {author} {\bibfnamefont {W.~E.}\ \bibnamefont {Pickett}},\ }\bibfield  {title} {\bibinfo {title} {Colloquium: Room temperature superconductivity: The roles of theory and materials design},\ }\href {https://doi.org/10.1103/RevModPhys.95.021001} {\bibfield  {journal} {\bibinfo  {journal} {Rev. Mod. Phys.}\ }\textbf {\bibinfo {volume} {95}},\ \bibinfo {pages} {021001} (\bibinfo {year} {2023})}\BibitemShut {NoStop}%
\bibitem [{\citenamefont {Stanev}\ \emph {et~al.}(2018)\citenamefont {Stanev}, \citenamefont {Oses}, \citenamefont {Kusne}, \citenamefont {Rodriguez}, \citenamefont {Paglione}, \citenamefont {Curtarolo},\ and\ \citenamefont {Takeuchi}}]{Stanev2018}%
  \BibitemOpen
  \bibfield  {author} {\bibinfo {author} {\bibfnamefont {V.}~\bibnamefont {Stanev}}, \bibinfo {author} {\bibfnamefont {C.}~\bibnamefont {Oses}}, \bibinfo {author} {\bibfnamefont {A.~G.}\ \bibnamefont {Kusne}}, \bibinfo {author} {\bibfnamefont {E.}~\bibnamefont {Rodriguez}}, \bibinfo {author} {\bibfnamefont {J.}~\bibnamefont {Paglione}}, \bibinfo {author} {\bibfnamefont {S.}~\bibnamefont {Curtarolo}},\ and\ \bibinfo {author} {\bibfnamefont {I.}~\bibnamefont {Takeuchi}},\ }\bibfield  {title} {\bibinfo {title} {{Machine learning modeling of superconducting critical temperature}},\ }\href {https://doi.org/10.1038/s41524-018-0085-8} {\bibfield  {journal} {\bibinfo  {journal} {npj Computational Materials}\ }\textbf {\bibinfo {volume} {4}},\ \bibinfo {pages} {29} (\bibinfo {year} {2018})}\BibitemShut {NoStop}%
\bibitem [{\citenamefont {Meredig}\ \emph {et~al.}(2018)\citenamefont {Meredig}, \citenamefont {Antono}, \citenamefont {Church}, \citenamefont {Hutchinson}, \citenamefont {Ling}, \citenamefont {Paradiso}, \citenamefont {Blaiszik}, \citenamefont {Foster}, \citenamefont {Gibbons}, \citenamefont {Hattrick-Simpers}, \citenamefont {Mehta},\ and\ \citenamefont {Ward}}]{Meredig2018}%
  \BibitemOpen
  \bibfield  {author} {\bibinfo {author} {\bibfnamefont {B.}~\bibnamefont {Meredig}}, \bibinfo {author} {\bibfnamefont {E.}~\bibnamefont {Antono}}, \bibinfo {author} {\bibfnamefont {C.}~\bibnamefont {Church}}, \bibinfo {author} {\bibfnamefont {M.}~\bibnamefont {Hutchinson}}, \bibinfo {author} {\bibfnamefont {J.}~\bibnamefont {Ling}}, \bibinfo {author} {\bibfnamefont {S.}~\bibnamefont {Paradiso}}, \bibinfo {author} {\bibfnamefont {B.}~\bibnamefont {Blaiszik}}, \bibinfo {author} {\bibfnamefont {I.}~\bibnamefont {Foster}}, \bibinfo {author} {\bibfnamefont {B.}~\bibnamefont {Gibbons}}, \bibinfo {author} {\bibfnamefont {J.}~\bibnamefont {Hattrick-Simpers}}, \bibinfo {author} {\bibfnamefont {A.}~\bibnamefont {Mehta}},\ and\ \bibinfo {author} {\bibfnamefont {L.}~\bibnamefont {Ward}},\ }\bibfield  {title} {\bibinfo {title} {{Can machine learning identify the next high-temperature superconductor? Examining extrapolation performance for materials discovery}},\ }\href {https://doi.org/10.1039/C8ME00012C} {\bibfield
  {journal} {\bibinfo  {journal} {Mol. Syst. Des. Eng.}\ }\textbf {\bibinfo {volume} {3}},\ \bibinfo {pages} {819} (\bibinfo {year} {2018})}\BibitemShut {NoStop}%
\bibitem [{\citenamefont {Hutcheon}\ \emph {et~al.}(2020)\citenamefont {Hutcheon}, \citenamefont {Shipley},\ and\ \citenamefont {Needs}}]{Hutcheon2020}%
  \BibitemOpen
  \bibfield  {author} {\bibinfo {author} {\bibfnamefont {M.~J.}\ \bibnamefont {Hutcheon}}, \bibinfo {author} {\bibfnamefont {A.~M.}\ \bibnamefont {Shipley}},\ and\ \bibinfo {author} {\bibfnamefont {R.~J.}\ \bibnamefont {Needs}},\ }\bibfield  {title} {\bibinfo {title} {Predicting novel superconducting hydrides using machine learning approaches},\ }\href {https://doi.org/10.1103/PhysRevB.101.144505} {\bibfield  {journal} {\bibinfo  {journal} {Phys. Rev. B}\ }\textbf {\bibinfo {volume} {101}},\ \bibinfo {pages} {144505} (\bibinfo {year} {2020})}\BibitemShut {NoStop}%
\bibitem [{\citenamefont {Xie}\ \emph {et~al.}(2022)\citenamefont {Xie}, \citenamefont {Quan}, \citenamefont {Hire}, \citenamefont {Deng}, \citenamefont {DeStefano}, \citenamefont {Salinas}, \citenamefont {Shah}, \citenamefont {Fanfarillo}, \citenamefont {Lim}, \citenamefont {Kim}, \citenamefont {Stewart}, \citenamefont {Hamlin}, \citenamefont {Hirschfeld},\ and\ \citenamefont {Hennig}}]{Xie2022}%
  \BibitemOpen
  \bibfield  {author} {\bibinfo {author} {\bibfnamefont {S.~R.}\ \bibnamefont {Xie}}, \bibinfo {author} {\bibfnamefont {Y.}~\bibnamefont {Quan}}, \bibinfo {author} {\bibfnamefont {A.~C.}\ \bibnamefont {Hire}}, \bibinfo {author} {\bibfnamefont {B.}~\bibnamefont {Deng}}, \bibinfo {author} {\bibfnamefont {J.~M.}\ \bibnamefont {DeStefano}}, \bibinfo {author} {\bibfnamefont {I.}~\bibnamefont {Salinas}}, \bibinfo {author} {\bibfnamefont {U.~S.}\ \bibnamefont {Shah}}, \bibinfo {author} {\bibfnamefont {L.}~\bibnamefont {Fanfarillo}}, \bibinfo {author} {\bibfnamefont {J.}~\bibnamefont {Lim}}, \bibinfo {author} {\bibfnamefont {J.}~\bibnamefont {Kim}}, \bibinfo {author} {\bibfnamefont {G.~R.}\ \bibnamefont {Stewart}}, \bibinfo {author} {\bibfnamefont {J.~J.}\ \bibnamefont {Hamlin}}, \bibinfo {author} {\bibfnamefont {P.~J.}\ \bibnamefont {Hirschfeld}},\ and\ \bibinfo {author} {\bibfnamefont {R.~G.}\ \bibnamefont {Hennig}},\ }\bibfield  {title} {\bibinfo {title} {Machine learning of superconducting critical temperature
  from eliashberg theory},\ }\href {https://doi.org/10.1038/s41524-021-00666-7} {\bibfield  {journal} {\bibinfo  {journal} {npj Computational Materials}\ }\textbf {\bibinfo {volume} {8}},\ \bibinfo {pages} {14} (\bibinfo {year} {2022})}\BibitemShut {NoStop}%
\bibitem [{\citenamefont {Zhang}\ \emph {et~al.}(2022)\citenamefont {Zhang}, \citenamefont {Cui}, \citenamefont {Hutcheon}, \citenamefont {Shipley}, \citenamefont {Song}, \citenamefont {Du}, \citenamefont {Kresin}, \citenamefont {Duan}, \citenamefont {Pickard},\ and\ \citenamefont {Yao}}]{ZhangPickard:2022}%
  \BibitemOpen
  \bibfield  {author} {\bibinfo {author} {\bibfnamefont {Z.}~\bibnamefont {Zhang}}, \bibinfo {author} {\bibfnamefont {T.}~\bibnamefont {Cui}}, \bibinfo {author} {\bibfnamefont {M.~J.}\ \bibnamefont {Hutcheon}}, \bibinfo {author} {\bibfnamefont {A.~M.}\ \bibnamefont {Shipley}}, \bibinfo {author} {\bibfnamefont {H.}~\bibnamefont {Song}}, \bibinfo {author} {\bibfnamefont {M.}~\bibnamefont {Du}}, \bibinfo {author} {\bibfnamefont {V.~Z.}\ \bibnamefont {Kresin}}, \bibinfo {author} {\bibfnamefont {D.}~\bibnamefont {Duan}}, \bibinfo {author} {\bibfnamefont {C.~J.}\ \bibnamefont {Pickard}},\ and\ \bibinfo {author} {\bibfnamefont {Y.}~\bibnamefont {Yao}},\ }\bibfield  {title} {\bibinfo {title} {{Design Principles for High-Temperature Superconductors with a Hydrogen-Based Alloy Backbone at Moderate Pressure}},\ }\href {https://doi.org/10.1103/PhysRevLett.128.047001} {\bibfield  {journal} {\bibinfo  {journal} {Phys. Rev. Lett.}\ }\textbf {\bibinfo {volume} {128}},\ \bibinfo {pages} {047001} (\bibinfo {year}
  {2022})}\BibitemShut {NoStop}%
\bibitem [{\citenamefont {Choudhary}\ and\ \citenamefont {Garrity}(2022)}]{Choudhary2022}%
  \BibitemOpen
  \bibfield  {author} {\bibinfo {author} {\bibfnamefont {K.}~\bibnamefont {Choudhary}}\ and\ \bibinfo {author} {\bibfnamefont {K.}~\bibnamefont {Garrity}},\ }\bibfield  {title} {\bibinfo {title} {{Designing high-$T_c$ superconductors with BCS-inspired screening, density functional theory, and deep-learning}},\ }\href {https://doi.org/10.1038/s41524-022-00933-1} {\bibfield  {journal} {\bibinfo  {journal} {npj Computational Materials}\ }\textbf {\bibinfo {volume} {8}},\ \bibinfo {pages} {244} (\bibinfo {year} {2022})}\BibitemShut {NoStop}%
\bibitem [{\citenamefont {Wines}\ \emph {et~al.}(2023)\citenamefont {Wines}, \citenamefont {Choudhary}, \citenamefont {Biacchi}, \citenamefont {Garrity},\ and\ \citenamefont {Tavazza}}]{Wines2023}%
  \BibitemOpen
  \bibfield  {author} {\bibinfo {author} {\bibfnamefont {D.}~\bibnamefont {Wines}}, \bibinfo {author} {\bibfnamefont {K.}~\bibnamefont {Choudhary}}, \bibinfo {author} {\bibfnamefont {A.~J.}\ \bibnamefont {Biacchi}}, \bibinfo {author} {\bibfnamefont {K.~F.}\ \bibnamefont {Garrity}},\ and\ \bibinfo {author} {\bibfnamefont {F.}~\bibnamefont {Tavazza}},\ }\bibfield  {title} {\bibinfo {title} {{High-Throughput DFT-Based Discovery of Next Generation Two-Dimensional (2D) Superconductors}},\ }\href {https://doi.org/10.1021/acs.nanolett.2c04420} {\bibfield  {journal} {\bibinfo  {journal} {Nano Letters}\ }\textbf {\bibinfo {volume} {23}},\ \bibinfo {pages} {969} (\bibinfo {year} {2023})}\BibitemShut {NoStop}%
\bibitem [{\citenamefont {Sommer}\ \emph {et~al.}(2023)\citenamefont {Sommer}, \citenamefont {Willa}, \citenamefont {Schmalian},\ and\ \citenamefont {Friederich}}]{Sommer2022_3DSC}%
  \BibitemOpen
  \bibfield  {author} {\bibinfo {author} {\bibfnamefont {T.}~\bibnamefont {Sommer}}, \bibinfo {author} {\bibfnamefont {R.}~\bibnamefont {Willa}}, \bibinfo {author} {\bibfnamefont {J.}~\bibnamefont {Schmalian}},\ and\ \bibinfo {author} {\bibfnamefont {P.}~\bibnamefont {Friederich}},\ }\bibfield  {title} {\bibinfo {title} {{3DSC - a dataset of superconductors including crystal structures}},\ }\href {https://doi.org/10.1038/s41597-023-02721-y} {\bibfield  {journal} {\bibinfo  {journal} {Scientific Data}\ }\textbf {\bibinfo {volume} {10}},\ \bibinfo {pages} {816} (\bibinfo {year} {2023})}\BibitemShut {NoStop}%
\bibitem [{\citenamefont {Gibson}\ \emph {et~al.}(2025{\natexlab{a}})\citenamefont {Gibson}, \citenamefont {Hire}, \citenamefont {Dee}, \citenamefont {Barrera}, \citenamefont {Geisler}, \citenamefont {Hirschfeld},\ and\ \citenamefont {Hennig}}]{betenet}%
  \BibitemOpen
  \bibfield  {author} {\bibinfo {author} {\bibfnamefont {J.~B.}\ \bibnamefont {Gibson}}, \bibinfo {author} {\bibfnamefont {A.~C.}\ \bibnamefont {Hire}}, \bibinfo {author} {\bibfnamefont {P.~M.}\ \bibnamefont {Dee}}, \bibinfo {author} {\bibfnamefont {O.}~\bibnamefont {Barrera}}, \bibinfo {author} {\bibfnamefont {B.}~\bibnamefont {Geisler}}, \bibinfo {author} {\bibfnamefont {P.~J.}\ \bibnamefont {Hirschfeld}},\ and\ \bibinfo {author} {\bibfnamefont {R.~G.}\ \bibnamefont {Hennig}},\ }\bibfield  {title} {\bibinfo {title} {Accelerating superconductor discovery through tempered deep learning of the electron-phonon spectral function},\ }\href {https://doi.org/10.1038/s41524-024-01475-4} {\bibfield  {journal} {\bibinfo  {journal} {npj Computational Materials}\ }\textbf {\bibinfo {volume} {11}},\ \bibinfo {pages} {7} (\bibinfo {year} {2025}{\natexlab{a}})}\BibitemShut {NoStop}%
\bibitem [{\citenamefont {Gibson}\ \emph {et~al.}(2025{\natexlab{b}})\citenamefont {Gibson}, \citenamefont {Hire}, \citenamefont {Prakash}, \citenamefont {Dee}, \citenamefont {Geisler}, \citenamefont {Kim}, \citenamefont {Li}, \citenamefont {Hamlin}, \citenamefont {Stewart}, \citenamefont {Hirschfeld},\ and\ \citenamefont {Hennig}}]{beenet}%
  \BibitemOpen
  \bibfield  {author} {\bibinfo {author} {\bibfnamefont {J.~B.}\ \bibnamefont {Gibson}}, \bibinfo {author} {\bibfnamefont {A.~C.}\ \bibnamefont {Hire}}, \bibinfo {author} {\bibfnamefont {P.}~\bibnamefont {Prakash}}, \bibinfo {author} {\bibfnamefont {P.~M.}\ \bibnamefont {Dee}}, \bibinfo {author} {\bibfnamefont {B.}~\bibnamefont {Geisler}}, \bibinfo {author} {\bibfnamefont {J.~S.}\ \bibnamefont {Kim}}, \bibinfo {author} {\bibfnamefont {Z.}~\bibnamefont {Li}}, \bibinfo {author} {\bibfnamefont {J.~J.}\ \bibnamefont {Hamlin}}, \bibinfo {author} {\bibfnamefont {G.~R.}\ \bibnamefont {Stewart}}, \bibinfo {author} {\bibfnamefont {P.~J.}\ \bibnamefont {Hirschfeld}},\ and\ \bibinfo {author} {\bibfnamefont {R.~G.}\ \bibnamefont {Hennig}},\ }\href {https://arxiv.org/abs/2503.20005} {\bibinfo {title} {Developing a complete {AI}-accelerated workflow for superconductor discovery}} (\bibinfo {year} {2025}{\natexlab{b}}),\ \Eprint {https://arxiv.org/abs/2503.20005} {arXiv:2503.20005 [cond-mat.supr-con]} \BibitemShut {NoStop}%
\bibitem [{\citenamefont {Prakash}\ \emph {et~al.}(2025)\citenamefont {Prakash}, \citenamefont {Gibson}, \citenamefont {Li}, \citenamefont {Gianluca}, \citenamefont {Esquivel}, \citenamefont {Fuemmeler}, \citenamefont {Geisler}, \citenamefont {Kim}, \citenamefont {Roitberg}, \citenamefont {Tadmor}, \citenamefont {Liu}, \citenamefont {Martiniani}, \citenamefont {Stewart}, \citenamefont {Hamlin}, \citenamefont {Hirschfeld},\ and\ \citenamefont {Hennig}}]{prakash-guideddiffusion}%
  \BibitemOpen
  \bibfield  {author} {\bibinfo {author} {\bibfnamefont {P.}~\bibnamefont {Prakash}}, \bibinfo {author} {\bibfnamefont {J.~B.}\ \bibnamefont {Gibson}}, \bibinfo {author} {\bibfnamefont {Z.}~\bibnamefont {Li}}, \bibinfo {author} {\bibfnamefont {G.~D.}\ \bibnamefont {Gianluca}}, \bibinfo {author} {\bibfnamefont {J.}~\bibnamefont {Esquivel}}, \bibinfo {author} {\bibfnamefont {E.}~\bibnamefont {Fuemmeler}}, \bibinfo {author} {\bibfnamefont {B.}~\bibnamefont {Geisler}}, \bibinfo {author} {\bibfnamefont {J.~S.}\ \bibnamefont {Kim}}, \bibinfo {author} {\bibfnamefont {A.}~\bibnamefont {Roitberg}}, \bibinfo {author} {\bibfnamefont {E.~B.}\ \bibnamefont {Tadmor}}, \bibinfo {author} {\bibfnamefont {M.}~\bibnamefont {Liu}}, \bibinfo {author} {\bibfnamefont {S.}~\bibnamefont {Martiniani}}, \bibinfo {author} {\bibfnamefont {G.~R.}\ \bibnamefont {Stewart}}, \bibinfo {author} {\bibfnamefont {J.~J.}\ \bibnamefont {Hamlin}}, \bibinfo {author} {\bibfnamefont {P.~J.}\ \bibnamefont {Hirschfeld}},\ and\ \bibinfo {author}
  {\bibfnamefont {R.~G.}\ \bibnamefont {Hennig}},\ }\href {https://arxiv.org/abs/2509.25186} {\bibinfo {title} {Guided diffusion for the discovery of new superconductors}} (\bibinfo {year} {2025}),\ \Eprint {https://arxiv.org/abs/2509.25186} {arXiv:2509.25186 [cond-mat.supr-con]} \BibitemShut {NoStop}%
\bibitem [{\citenamefont {Larbalestier}\ \emph {et~al.}(2001)\citenamefont {Larbalestier}, \citenamefont {Gurevich}, \citenamefont {Feldmann},\ and\ \citenamefont {Polyanskii}}]{Larbalestier2001}%
  \BibitemOpen
  \bibfield  {author} {\bibinfo {author} {\bibfnamefont {D.}~\bibnamefont {Larbalestier}}, \bibinfo {author} {\bibfnamefont {A.}~\bibnamefont {Gurevich}}, \bibinfo {author} {\bibfnamefont {D.~M.}\ \bibnamefont {Feldmann}},\ and\ \bibinfo {author} {\bibfnamefont {A.}~\bibnamefont {Polyanskii}},\ }\bibfield  {title} {\bibinfo {title} {{High-$T_{c}$ superconducting materials for electric power applications}},\ }\href {https://doi.org/10.1038/35104654} {\bibfield  {journal} {\bibinfo  {journal} {Nature}\ }\textbf {\bibinfo {volume} {414}},\ \bibinfo {pages} {368} (\bibinfo {year} {2001})}\BibitemShut {NoStop}%
\bibitem [{\citenamefont {Blatter}\ \emph {et~al.}(1994)\citenamefont {Blatter}, \citenamefont {Feigel'man}, \citenamefont {Geshkenbein}, \citenamefont {Larkin},\ and\ \citenamefont {Vinokur}}]{Vortices:1994}%
  \BibitemOpen
  \bibfield  {author} {\bibinfo {author} {\bibfnamefont {G.}~\bibnamefont {Blatter}}, \bibinfo {author} {\bibfnamefont {M.~V.}\ \bibnamefont {Feigel'man}}, \bibinfo {author} {\bibfnamefont {V.~B.}\ \bibnamefont {Geshkenbein}}, \bibinfo {author} {\bibfnamefont {A.~I.}\ \bibnamefont {Larkin}},\ and\ \bibinfo {author} {\bibfnamefont {V.~M.}\ \bibnamefont {Vinokur}},\ }\bibfield  {title} {\bibinfo {title} {Vortices in high-temperature superconductors},\ }\href {https://doi.org/10.1103/RevModPhys.66.1125} {\bibfield  {journal} {\bibinfo  {journal} {Rev. Mod. Phys.}\ }\textbf {\bibinfo {volume} {66}},\ \bibinfo {pages} {1125} (\bibinfo {year} {1994})}\BibitemShut {NoStop}%
\bibitem [{\citenamefont {Choi}\ \emph {et~al.}(2002)\citenamefont {Choi}, \citenamefont {Roundy}, \citenamefont {Sun}, \citenamefont {Cohen},\ and\ \citenamefont {Louie}}]{Choi2002}%
  \BibitemOpen
  \bibfield  {author} {\bibinfo {author} {\bibfnamefont {H.~J.}\ \bibnamefont {Choi}}, \bibinfo {author} {\bibfnamefont {D.}~\bibnamefont {Roundy}}, \bibinfo {author} {\bibfnamefont {H.}~\bibnamefont {Sun}}, \bibinfo {author} {\bibfnamefont {M.~L.}\ \bibnamefont {Cohen}},\ and\ \bibinfo {author} {\bibfnamefont {S.~G.}\ \bibnamefont {Louie}},\ }\bibfield  {title} {\bibinfo {title} {{First-principles calculation of the superconducting transition in MgB$_2$ within the anisotropic Eliashberg formalism}},\ }\href {https://doi.org/10.1103/PhysRevB.66.020513} {\bibfield  {journal} {\bibinfo  {journal} {Phys. Rev. B}\ }\textbf {\bibinfo {volume} {66}},\ \bibinfo {pages} {020513} (\bibinfo {year} {2002})}\BibitemShut {NoStop}%
\bibitem [{\citenamefont {Margine}\ and\ \citenamefont {Giustino}(2013)}]{Margine2013}%
  \BibitemOpen
  \bibfield  {author} {\bibinfo {author} {\bibfnamefont {E.~R.}\ \bibnamefont {Margine}}\ and\ \bibinfo {author} {\bibfnamefont {F.}~\bibnamefont {Giustino}},\ }\bibfield  {title} {\bibinfo {title} {{Anisotropic Migdal-Eliashberg theory using Wannier functions}},\ }\href {https://doi.org/10.1103/PhysRevB.87.024505} {\bibfield  {journal} {\bibinfo  {journal} {Phys. Rev. B}\ }\textbf {\bibinfo {volume} {87}},\ \bibinfo {pages} {024505} (\bibinfo {year} {2013})}\BibitemShut {NoStop}%
\bibitem [{\citenamefont {{National Academies of Sciences, Engineering, and Medicine}}(2024)}]{NAS_HighField2024}%
  \BibitemOpen
  \bibfield  {author} {\bibinfo {author} {\bibnamefont {{National Academies of Sciences, Engineering, and Medicine}}},\ }\href {https://doi.org/10.17226/27830} {\emph {\bibinfo {title} {The Current Status and Future Direction of High-Magnetic-Field Science and Technology in the United States}}}\ (\bibinfo  {publisher} {The National Academies Press},\ \bibinfo {address} {Washington, DC},\ \bibinfo {year} {2024})\BibitemShut {NoStop}%
\bibitem [{\citenamefont {Kohn}\ and\ \citenamefont {Sham}(1965)}]{KoSh65}%
  \BibitemOpen
  \bibfield  {author} {\bibinfo {author} {\bibfnamefont {W.}~\bibnamefont {Kohn}}\ and\ \bibinfo {author} {\bibfnamefont {L.~J.}\ \bibnamefont {Sham}},\ }\bibfield  {title} {\bibinfo {title} {Self-consistent equations including exchange and correlation effects},\ }\href {https://doi.org/10.1103/PhysRev.140.A1133} {\bibfield  {journal} {\bibinfo  {journal} {Phys. Rev.}\ }\textbf {\bibinfo {volume} {140}},\ \bibinfo {pages} {A1133} (\bibinfo {year} {1965})}\BibitemShut {NoStop}%
\bibitem [{\citenamefont {Eliashberg}(1960)}]{EliashbergInteractionBetweenElAndLatticeVibrInASC1960}%
  \BibitemOpen
  \bibfield  {author} {\bibinfo {author} {\bibfnamefont {G.~M.}\ \bibnamefont {Eliashberg}},\ }\bibfield  {title} {\bibinfo {title} {{Interaction between electrons and lattice vibrations in a superconductor}},\ }\href@noop {} {\bibfield  {journal} {\bibinfo  {journal} {Sov. Phys. JETP}\ }\textbf {\bibinfo {volume} {11}},\ \bibinfo {pages} {696} (\bibinfo {year} {1960})}\BibitemShut {NoStop}%
\bibitem [{\citenamefont {Carbotte}(1990)}]{Carbotte1990RMP}%
  \BibitemOpen
  \bibfield  {author} {\bibinfo {author} {\bibfnamefont {J.~P.}\ \bibnamefont {Carbotte}},\ }\bibfield  {title} {\bibinfo {title} {Properties of boson-exchange superconductors},\ }\href {https://doi.org/10.1103/RevModPhys.62.1027} {\bibfield  {journal} {\bibinfo  {journal} {Rev. Mod. Phys.}\ }\textbf {\bibinfo {volume} {62}},\ \bibinfo {pages} {1027} (\bibinfo {year} {1990})}\BibitemShut {NoStop}%
\bibitem [{\citenamefont {Cerqueira}\ \emph {et~al.}(2023)\citenamefont {Cerqueira}, \citenamefont {Sanna},\ and\ \citenamefont {Marques}}]{Cerqueira2023}%
  \BibitemOpen
  \bibfield  {author} {\bibinfo {author} {\bibfnamefont {T.~F.~T.}\ \bibnamefont {Cerqueira}}, \bibinfo {author} {\bibfnamefont {A.}~\bibnamefont {Sanna}},\ and\ \bibinfo {author} {\bibfnamefont {M.~A.~L.}\ \bibnamefont {Marques}},\ }\bibfield  {title} {\bibinfo {title} {Sampling the materials space for conventional superconducting compounds},\ }\bibfield  {journal} {\bibinfo  {journal} {Advanced Materials}\ }\href {https://doi.org/10.1002/adma.202307085} {10.1002/adma.202307085} (\bibinfo {year} {2023})\BibitemShut {NoStop}%
\bibitem [{\citenamefont {Ponc\'e}\ \emph {et~al.}(2016)\citenamefont {Ponc\'e}, \citenamefont {Margine}, \citenamefont {Verdi},\ and\ \citenamefont {Giustino}}]{EPW-1}%
  \BibitemOpen
  \bibfield  {author} {\bibinfo {author} {\bibfnamefont {S.}~\bibnamefont {Ponc\'e}}, \bibinfo {author} {\bibfnamefont {E.}~\bibnamefont {Margine}}, \bibinfo {author} {\bibfnamefont {C.}~\bibnamefont {Verdi}},\ and\ \bibinfo {author} {\bibfnamefont {F.}~\bibnamefont {Giustino}},\ }\bibfield  {title} {\bibinfo {title} {{EPW}: Electron-phonon coupling, transport and superconducting properties using maximally localized {Wannier} functions},\ }\href {https://doi.org/https://doi.org/10.1016/j.cpc.2016.07.028} {\bibfield  {journal} {\bibinfo  {journal} {Computer Physics Communications}\ }\textbf {\bibinfo {volume} {209}},\ \bibinfo {pages} {116 } (\bibinfo {year} {2016})}\BibitemShut {NoStop}%
\bibitem [{\citenamefont {Giustino}\ \emph {et~al.}(2007)\citenamefont {Giustino}, \citenamefont {Cohen},\ and\ \citenamefont {Louie}}]{EPW-2}%
  \BibitemOpen
  \bibfield  {author} {\bibinfo {author} {\bibfnamefont {F.}~\bibnamefont {Giustino}}, \bibinfo {author} {\bibfnamefont {M.~L.}\ \bibnamefont {Cohen}},\ and\ \bibinfo {author} {\bibfnamefont {S.~G.}\ \bibnamefont {Louie}},\ }\bibfield  {title} {\bibinfo {title} {Electron-phonon interaction using {Wannier} functions},\ }\href {https://doi.org/10.1103/PhysRevB.76.165108} {\bibfield  {journal} {\bibinfo  {journal} {Phys. Rev. B}\ }\textbf {\bibinfo {volume} {76}},\ \bibinfo {pages} {165108} (\bibinfo {year} {2007})}\BibitemShut {NoStop}%
\bibitem [{\citenamefont {Kresse}\ and\ \citenamefont {Furthm{\"{u}}ller}(1996{\natexlab{a}})}]{Kresse1996}%
  \BibitemOpen
  \bibfield  {author} {\bibinfo {author} {\bibfnamefont {G.}~\bibnamefont {Kresse}}\ and\ \bibinfo {author} {\bibfnamefont {J.}~\bibnamefont {Furthm{\"{u}}ller}},\ }\bibfield  {title} {\bibinfo {title} {{Efficient iterative schemes for ab initio total-energy calculations using a plane-wave basis set}},\ }\href {https://doi.org/10.1103/PhysRevB.54.11169} {\bibfield  {journal} {\bibinfo  {journal} {Phys. Rev. B}\ }\textbf {\bibinfo {volume} {54}},\ \bibinfo {pages} {11169} (\bibinfo {year} {1996}{\natexlab{a}})}\BibitemShut {NoStop}%
\bibitem [{\citenamefont {Kresse}\ and\ \citenamefont {Furthm{\"{u}}ller}(1996{\natexlab{b}})}]{Kresse1996b}%
  \BibitemOpen
  \bibfield  {author} {\bibinfo {author} {\bibfnamefont {G.}~\bibnamefont {Kresse}}\ and\ \bibinfo {author} {\bibfnamefont {J.}~\bibnamefont {Furthm{\"{u}}ller}},\ }\bibfield  {title} {\bibinfo {title} {Efficiency of ab-initio total energy calculations for metals and semiconductors using a plane-wave basis set},\ }\href {https://doi.org/https://doi.org/10.1016/0927-0256(96)00008-0} {\bibfield  {journal} {\bibinfo  {journal} {Computational Materials Science}\ }\textbf {\bibinfo {volume} {6}},\ \bibinfo {pages} {15} (\bibinfo {year} {1996}{\natexlab{b}})}\BibitemShut {NoStop}%
\bibitem [{\citenamefont {Kresse}\ and\ \citenamefont {Joubert}(1999)}]{Kresse1999}%
  \BibitemOpen
  \bibfield  {author} {\bibinfo {author} {\bibfnamefont {G.}~\bibnamefont {Kresse}}\ and\ \bibinfo {author} {\bibfnamefont {D.}~\bibnamefont {Joubert}},\ }\bibfield  {title} {\bibinfo {title} {From ultrasoft pseudopotentials to the projector augmented-wave method},\ }\href {https://doi.org/10.1103/PhysRevB.59.1758} {\bibfield  {journal} {\bibinfo  {journal} {Phys. Rev. B}\ }\textbf {\bibinfo {volume} {59}},\ \bibinfo {pages} {1758} (\bibinfo {year} {1999})}\BibitemShut {NoStop}%
\bibitem [{\citenamefont {Whitmore}\ and\ \citenamefont {Carbotte}(1981)}]{Whitmore1981}%
  \BibitemOpen
  \bibfield  {author} {\bibinfo {author} {\bibfnamefont {M.~D.}\ \bibnamefont {Whitmore}}\ and\ \bibinfo {author} {\bibfnamefont {J.~P.}\ \bibnamefont {Carbotte}},\ }\bibfield  {title} {\bibinfo {title} {Effects of impurities on anisotropic superconductors with repulsive average interaction},\ }\href {https://doi.org/10.1088/0305-4608/11/12/011} {\bibfield  {journal} {\bibinfo  {journal} {Journal of Physics F: Metal Physics}\ }\textbf {\bibinfo {volume} {11}},\ \bibinfo {pages} {2585} (\bibinfo {year} {1981})}\BibitemShut {NoStop}%
\bibitem [{\citenamefont {Tinkham}(2004)}]{Tinkham-SC}%
  \BibitemOpen
  \bibfield  {author} {\bibinfo {author} {\bibfnamefont {M.}~\bibnamefont {Tinkham}},\ }\href {https://books.google.com/books?id=VpUk3NfwDIkC} {\emph {\bibinfo {title} {Introduction to Superconductivity}}},\ Dover Books on Physics Series\ (\bibinfo  {publisher} {Dover Publications},\ \bibinfo {year} {2004})\BibitemShut {NoStop}%
\bibitem [{\citenamefont {Brandt}(2011)}]{Brandt-Hc1:2011}%
  \BibitemOpen
  \bibfield  {author} {\bibinfo {author} {\bibfnamefont {E.~H.}\ \bibnamefont {Brandt}},\ }\bibfield  {title} {\bibinfo {title} {The vortex lattice in {Type-II} superconductors: {Ideal} or distorted, in bulk and films},\ }\href {https://doi.org/https://doi.org/10.1002/pssb.201147095} {\bibfield  {journal} {\bibinfo  {journal} {physica status solidi (b)}\ }\textbf {\bibinfo {volume} {248}},\ \bibinfo {pages} {2305} (\bibinfo {year} {2011})}\BibitemShut {NoStop}%
\bibitem [{\citenamefont {Mitrovi{\'{c}}}\ \emph {et~al.}(1984)\citenamefont {Mitrovi{\'{c}}}, \citenamefont {Zarate},\ and\ \citenamefont {Carbotte}}]{Mitrovic1984}%
  \BibitemOpen
  \bibfield  {author} {\bibinfo {author} {\bibfnamefont {B.}~\bibnamefont {Mitrovi{\'{c}}}}, \bibinfo {author} {\bibfnamefont {H.~G.}\ \bibnamefont {Zarate}},\ and\ \bibinfo {author} {\bibfnamefont {J.~P.}\ \bibnamefont {Carbotte}},\ }\bibfield  {title} {\bibinfo {title} {{The ratio $2\Delta_0 / k_\text{B}T_\text{c}$ within Eliashberg theory}},\ }\href {https://doi.org/10.1103/PhysRevB.29.184} {\bibfield  {journal} {\bibinfo  {journal} {Phys. Rev. B}\ }\textbf {\bibinfo {volume} {29}},\ \bibinfo {pages} {184} (\bibinfo {year} {1984})}\BibitemShut {NoStop}%
\bibitem [{\citenamefont {Roberts}(1976)}]{Roberts-SC:1976}%
  \BibitemOpen
  \bibfield  {author} {\bibinfo {author} {\bibfnamefont {B.~W.}\ \bibnamefont {Roberts}},\ }\bibfield  {title} {\bibinfo {title} {Survey of superconductive materials and critical evaluation of selected properties},\ }\href {https://doi.org/10.1063/1.555540} {\bibfield  {journal} {\bibinfo  {journal} {Journal of Physical and Chemical Reference Data}\ }\textbf {\bibinfo {volume} {5}},\ \bibinfo {pages} {581} (\bibinfo {year} {1976})}\BibitemShut {NoStop}%
\bibitem [{\citenamefont {Zhu}\ \emph {et~al.}(2022)\citenamefont {Zhu}, \citenamefont {Zhang}, \citenamefont {Li}, \citenamefont {Li}, \citenamefont {Duan},\ and\ \citenamefont {Wen}}]{Cr3Ru:2022}%
  \BibitemOpen
  \bibfield  {author} {\bibinfo {author} {\bibfnamefont {Z.}~\bibnamefont {Zhu}}, \bibinfo {author} {\bibfnamefont {Y.-J.}\ \bibnamefont {Zhang}}, \bibinfo {author} {\bibfnamefont {Y.}~\bibnamefont {Li}}, \bibinfo {author} {\bibfnamefont {Q.}~\bibnamefont {Li}}, \bibinfo {author} {\bibfnamefont {W.}~\bibnamefont {Duan}},\ and\ \bibinfo {author} {\bibfnamefont {H.-H.}\ \bibnamefont {Wen}},\ }\bibfield  {title} {\bibinfo {title} {Comparative studies on superconductivity in {Cr$_3$Ru} compounds with bcc and {A15} structures},\ }\href {https://doi.org/10.1088/1361-648X/ac9501} {\bibfield  {journal} {\bibinfo  {journal} {Journal of Physics: Condensed Matter}\ }\textbf {\bibinfo {volume} {34}},\ \bibinfo {pages} {475602} (\bibinfo {year} {2022})}\BibitemShut {NoStop}%
\bibitem [{\citenamefont {Eisenstein}(1954)}]{Eisenstein:1954}%
  \BibitemOpen
  \bibfield  {author} {\bibinfo {author} {\bibfnamefont {J.}~\bibnamefont {Eisenstein}},\ }\bibfield  {title} {\bibinfo {title} {Superconducting elements},\ }\href {https://doi.org/10.1103/RevModPhys.26.277} {\bibfield  {journal} {\bibinfo  {journal} {Rev. Mod. Phys.}\ }\textbf {\bibinfo {volume} {26}},\ \bibinfo {pages} {277} (\bibinfo {year} {1954})}\BibitemShut {NoStop}%
\bibitem [{\citenamefont {Matthias}\ \emph {et~al.}(1963)\citenamefont {Matthias}, \citenamefont {Geballe},\ and\ \citenamefont {Compton}}]{MatthiasGeballeCompton:1963}%
  \BibitemOpen
  \bibfield  {author} {\bibinfo {author} {\bibfnamefont {B.~T.}\ \bibnamefont {Matthias}}, \bibinfo {author} {\bibfnamefont {T.~H.}\ \bibnamefont {Geballe}},\ and\ \bibinfo {author} {\bibfnamefont {V.~B.}\ \bibnamefont {Compton}},\ }\bibfield  {title} {\bibinfo {title} {Superconductivity},\ }\href {https://doi.org/10.1103/RevModPhys.35.1} {\bibfield  {journal} {\bibinfo  {journal} {Rev. Mod. Phys.}\ }\textbf {\bibinfo {volume} {35}},\ \bibinfo {pages} {1} (\bibinfo {year} {1963})}\BibitemShut {NoStop}%
\bibitem [{\citenamefont {Schwochau}(2000)}]{Schwochau-Tc:2000}%
  \BibitemOpen
  \bibfield  {author} {\bibinfo {author} {\bibfnamefont {K.}~\bibnamefont {Schwochau}},\ }\href {https://books.google.com/books?id=BHjxH8q9iukC} {\emph {\bibinfo {title} {Technetium: {Chemistry} and Radiopharmaceutical Applications}}}\ (\bibinfo  {publisher} {Wiley},\ \bibinfo {year} {2000})\BibitemShut {NoStop}%
\bibitem [{\citenamefont {Zehetmayer}\ \emph {et~al.}(2002)\citenamefont {Zehetmayer}, \citenamefont {Eisterer}, \citenamefont {Jun}, \citenamefont {Kazakov}, \citenamefont {Karpinski}, \citenamefont {Wisniewski},\ and\ \citenamefont {Weber}}]{MgB2-Hc2:2002}%
  \BibitemOpen
  \bibfield  {author} {\bibinfo {author} {\bibfnamefont {M.}~\bibnamefont {Zehetmayer}}, \bibinfo {author} {\bibfnamefont {M.}~\bibnamefont {Eisterer}}, \bibinfo {author} {\bibfnamefont {J.}~\bibnamefont {Jun}}, \bibinfo {author} {\bibfnamefont {S.~M.}\ \bibnamefont {Kazakov}}, \bibinfo {author} {\bibfnamefont {J.}~\bibnamefont {Karpinski}}, \bibinfo {author} {\bibfnamefont {A.}~\bibnamefont {Wisniewski}},\ and\ \bibinfo {author} {\bibfnamefont {H.~W.}\ \bibnamefont {Weber}},\ }\bibfield  {title} {\bibinfo {title} {Mixed-state properties of superconducting {${\mathrm{MgB}}_{2}$} single crystals},\ }\href {https://doi.org/10.1103/PhysRevB.66.052505} {\bibfield  {journal} {\bibinfo  {journal} {Phys. Rev. B}\ }\textbf {\bibinfo {volume} {66}},\ \bibinfo {pages} {052505} (\bibinfo {year} {2002})}\BibitemShut {NoStop}%
\bibitem [{Kit(2007)}]{Kittel:2007}%
  \BibitemOpen
  \href {https://books.google.com/books?id=F9Qu5c_hUaUC} {\emph {\bibinfo {title} {Introduction to Solid State Physics, 7th Ed}}}\ (\bibinfo  {publisher} {Wiley India Pvt. Limited},\ \bibinfo {year} {2007})\BibitemShut {NoStop}%
\bibitem [{\citenamefont {Xie}\ \emph {et~al.}(2019)\citenamefont {Xie}, \citenamefont {Stewart}, \citenamefont {Hamlin}, \citenamefont {Hirschfeld},\ and\ \citenamefont {Hennig}}]{Xie2019}%
  \BibitemOpen
  \bibfield  {author} {\bibinfo {author} {\bibfnamefont {S.~R.}\ \bibnamefont {Xie}}, \bibinfo {author} {\bibfnamefont {G.~R.}\ \bibnamefont {Stewart}}, \bibinfo {author} {\bibfnamefont {J.~J.}\ \bibnamefont {Hamlin}}, \bibinfo {author} {\bibfnamefont {P.~J.}\ \bibnamefont {Hirschfeld}},\ and\ \bibinfo {author} {\bibfnamefont {R.~G.}\ \bibnamefont {Hennig}},\ }\bibfield  {title} {\bibinfo {title} {Functional form of the superconducting critical temperature from machine learning},\ }\href {https://doi.org/10.1103/PhysRevB.100.174513} {\bibfield  {journal} {\bibinfo  {journal} {Phys. Rev. B}\ }\textbf {\bibinfo {volume} {100}},\ \bibinfo {pages} {174513} (\bibinfo {year} {2019})}\BibitemShut {NoStop}%
\bibitem [{\citenamefont {Perdew}\ \emph {et~al.}(2008)\citenamefont {Perdew}, \citenamefont {Ruzsinszky}, \citenamefont {Csonka}, \citenamefont {Vydrov}, \citenamefont {Scuseria}, \citenamefont {Constantin}, \citenamefont {Zhou},\ and\ \citenamefont {Burke}}]{Perdew2008}%
  \BibitemOpen
  \bibfield  {author} {\bibinfo {author} {\bibfnamefont {J.~P.}\ \bibnamefont {Perdew}}, \bibinfo {author} {\bibfnamefont {A.}~\bibnamefont {Ruzsinszky}}, \bibinfo {author} {\bibfnamefont {G.~I.}\ \bibnamefont {Csonka}}, \bibinfo {author} {\bibfnamefont {O.~A.}\ \bibnamefont {Vydrov}}, \bibinfo {author} {\bibfnamefont {G.~E.}\ \bibnamefont {Scuseria}}, \bibinfo {author} {\bibfnamefont {L.~A.}\ \bibnamefont {Constantin}}, \bibinfo {author} {\bibfnamefont {X.}~\bibnamefont {Zhou}},\ and\ \bibinfo {author} {\bibfnamefont {K.}~\bibnamefont {Burke}},\ }\bibfield  {title} {\bibinfo {title} {{Restoring the Density-Gradient Expansion for Exchange in Solids and Surfaces}},\ }\href {https://doi.org/10.1103/PhysRevLett.100.136406} {\bibfield  {journal} {\bibinfo  {journal} {Phys. Rev. Lett.}\ }\textbf {\bibinfo {volume} {100}},\ \bibinfo {pages} {136406} (\bibinfo {year} {2008})}\BibitemShut {NoStop}%
\bibitem [{\citenamefont {Gall}(2016)}]{Gall-FermiVelocitiesMetals:16}%
  \BibitemOpen
  \bibfield  {author} {\bibinfo {author} {\bibfnamefont {D.}~\bibnamefont {Gall}},\ }\bibfield  {title} {\bibinfo {title} {Electron mean free path in elemental metals},\ }\href {https://doi.org/10.1063/1.4942216} {\bibfield  {journal} {\bibinfo  {journal} {Journal of Applied Physics}\ }\textbf {\bibinfo {volume} {119}},\ \bibinfo {pages} {085101} (\bibinfo {year} {2016})}\BibitemShut {NoStop}%
\bibitem [{\citenamefont {Allen}\ and\ \citenamefont {Dynes}(1975)}]{Allen-Dynes1975}%
  \BibitemOpen
  \bibfield  {author} {\bibinfo {author} {\bibfnamefont {P.~B.}\ \bibnamefont {Allen}}\ and\ \bibinfo {author} {\bibfnamefont {R.~C.}\ \bibnamefont {Dynes}},\ }\bibfield  {title} {\bibinfo {title} {Transition temperature of strong-coupled superconductors reanalyzed},\ }\href {https://doi.org/10.1103/PhysRevB.12.905} {\bibfield  {journal} {\bibinfo  {journal} {Phys. Rev. B}\ }\textbf {\bibinfo {volume} {12}},\ \bibinfo {pages} {905} (\bibinfo {year} {1975})}\BibitemShut {NoStop}%
\end{thebibliography}
\end{document}